\documentclass{jpp}

\usepackage[english]{babel}

\usepackage{amsmath}
\usepackage{amssymb}
\usepackage{graphicx}
\usepackage{mathtools}
\usepackage{physics}
\usepackage[colorlinks=true, allcolors=blue]{hyperref}
\usepackage[dvipsnames,table]{xcolor}
\usepackage{empheq}
\usepackage{subcaption}
\usepackage{todonotes}
\usepackage{placeins}
\usepackage{natbib}
\usepackage{tikz}
\usetikzlibrary{decorations.text}
\usepackage{rotating}
\usepackage[outline]{contour}
\usepackage{suffix}

\usepackage{textcomp}
\DeclareUnicodeCharacter{2009}{\,}

\newcommand*\bs[1]{\boldsymbol{#1}}

\renewcommand*\dd[2]{\frac{\partial #1}{\partial #2}}
\newcommand*\DD[2]{\frac{\mathrm{d} #1}{\mathrm{d} #2}}

\newcommand{\bhat}{\ensuremath{\boldsymbol{\hat{b}}}}

\newcommand{\alphafl}{{\ensuremath{{\alpha}}}}
\newcommand{\alphaflz}{{\ensuremath{{\alpha_{0}}}}}

\newcommand{\RGK}{\ensuremath{\bs{R}}}
\newcommand{\epsGK}{\ensuremath{\varepsilon}}
\newcommand{\muGK}{\ensuremath{\mu}}

\newcommand*\gyroavgR[1]{\left\langle #1 \right\rangle_{\RGK}}

\newcommand{\vmtildea}{\ensuremath{\tilde{\bs{v}}_{Ms}}}

\newcommand{\omegakappai}{\ensuremath{\omega_{\kappa,i}}}
\newcommand{\omeganablaBi}{\ensuremath{\omega_{\nabla B,i}}}

\renewcommand*\abs[1]{|#1|}

\DeclarePairedDelimiterX\MeijerM[3]{\lparen}{\rparen}%
{\begin{smallmatrix}#1 \\ #2\end{smallmatrix}\delimsize\vert\,#3}

\newcommand\MeijerG[8][]{%
  G^{\,#2,#3}_{#4,#5}\MeijerM[#1]{#6}{#7}{#8}}

\WithSuffix\newcommand\MeijerG*[7]{%
  G^{\,#1,#2}_{#3,#4}\MeijerM*{#5}{#6}{#7}}

\renewcommand{\parallel}{{\scriptstyle \|}}
\newcommand{\smallparallel}{{\scriptscriptstyle \|}}

\title{Geodesic extended modes in low magnetic shear tokamaks and stellarators}
\author{Richard Nies\aff{1,2} \corresp{\email{richard.nies@physics.ox.ac.uk}}, Felix Parra\aff{1,2}}

\affiliation{
\aff{1} Department of Astrophysical Sciences, Princeton University, Princeton, NJ 08543, USA 
\aff{2} Princeton Plasma Physics Laboratory, Princeton, NJ 08540, USA
}
\date{\today}

\shorttitle{Geodesic extended modes in low magnetic shear tokamaks and stellarators}
\shortauthor{R. Nies and F. Parra}

\begin{document}

\maketitle


\begin{abstract}
        Theories of ion-scale microinstabilities in tokamaks and stellarators typically assume that the passing electrons respond adiabatically due to their fast propagation speed. However, when the magnetic shear becomes sufficiently small, ion-scale modes can extend far along the magnetic field and the non-adiabatic response of passing electrons becomes important. We derive a theory of extended modes at low magnetic shear through a multiscale expansion of the gyrokinetic equation. The theory elucidates the physics of the geodesic extended mode, a new type of microinstability. The new mode couples the non-adiabatic physics of both electrons and ions, unlike extended modes at magnetic shear of order unity. The theory is validated against gyrokinetic simulations and the parameter dependences of the new mode are studied.
\end{abstract}

\section{Introduction}

In tokamaks and stellarators, the large density and temperature gradients between the hot dense core and the cold dilute edge drive microinstabilities on the scale of the particles' gyroradii. These microinstabilities give rise to turbulence which mixes the plasma, causing large heat fluxes limiting the energy confinement. The most virulent microinstabilities are typically on ion gyroradius scales, such as the ion temperature gradient (ITG) mode \citep{rudakov_instability_1961, coppi_instabilities_1967, horton_toroidal_1981, biglari_toroidal_1989, cowley_considerations_1991}. When considering ion-scale microinstabilities, the passing electron response is generally assumed to be adiabatic. This assumption is motivated by the fast propagation rate of electrons $\omega_{\parallel,e} \sim v_{Te}/R$ parallel to the magnetic field. The typical frequency of ion-scale modes $\omega \sim v_{Ti}/R$ is comparatively slow, as $v_{Te}/v_{Ti} \sim (m_i/m_e)^{1/2} \gg 1$. Here, $R$ is the major radius of the tokamak or stellarator, taken to be the typical system scale length, and $v_{T{s}} = (2 T_{s}/m_{s})^{1/2}$ is the thermal speed of particles of species ${s}$ with temperature $T_{s}$ and mass $m_{s}$. 

Notwithstanding the fast electron propagation, the non-adiabatic response of trapped electrons can become important even for ion-scale modes. Electrons trapped in regions of small magnetic field strength cannot respond adiabatically; this non-adiabatic trapped electron response can substantially modify the growth rate of ITG modes \citep[see e.g.][]{dominski_how_2015} and give rise to trapped electron modes (TEMs) \citep{kadomtsev_plasma_1967, rosenbluth_lowfrequency_1968}. Furthermore, the non-adiabatic response of passing electrons can also affect ion-scale modes. The non-adiabatic passing electron response typically leads to modes which extend far along the magnetic field line. Observations of such extended modes go back to early gyrokinetic simulations including kinetic electrons \citep{hallatschek_giant_2005} and they have since been the object of various numerical studies \citep{dominski_how_2015, belli_reversal_2019, ball_eliminating_2020, c_j_how_2020, c_j_effect_2021}. 

By performing a multiscale expansion of the gyrokinetic system of equations in $m_e/m_i \ll 1$, \cite{hardman_extended_2022} uncovered two types of electrostatic modes: the `small electron tail modes' and the `dominant electron tail modes', whose mode frequencies are determined by non-adiabatic ion physics and by non-adiabatic electron physics, respectively. In the expansion of \cite{hardman_extended_2022}, the magnetic shear $\hat s$, which describes the change in the field line pitch across flux surfaces, was assumed to be of order unity. For $\hat s \sim 1$, the non-adiabatic ion response vanishes for $\theta \gg 1$ due to finite Larmor radius (FLR) effects. Here, $\theta$ is the poloidal angle and will be used throughout this work as a coordinate parallel to the magnetic field lines. Because of the ion FLR stabilisation, the `small electron tail modes' are peaked at $\theta \sim 1$ and the non-adiabatic passing electron response only modifies the decay of the mode eigenfunction at $\theta \gg 1$. In contrast, the eigenfunctions of the `dominant electron tail modes' are extended in $\theta$ and the non-adiabatic ion response only causes a small change to the eigenfunction at $\theta \sim 1$.

In this work, we present a theory of extended modes for small values of the magnetic shear $\hat s \sim (m_e/m_i)^{1/2}$. With this ordering of $\hat s$, the ion FLR effects become substantial only for large $\theta \sim (m_i/m_e)^{1/2}$, the typical distance of electron propagation within a mode period $\omega^{-1}\sim R/v_{Ti}$. As a result, the mode frequency and eigenfunction are generally determined by non-adiabatic physics of both the electrons and the ions, unlike for the $\hat s \sim 1$ case. Notably, small values of the magnetic shear allow for a new type of microinstability. The new mode is prominent in the precise quasisymmetric stellarators of \cite{landreman_magnetic_2022} (see Section~\ref{sec:longtail_precise_QA}). The mode is also a salient feature of tokamak simulations at small magnetic shear, where it was called a `$k_z \approx 0 $' toroidal ITG mode \citep{volcokas_ultra_2023, volcokas_numerical_2024}. Henceforth, we call the new mode a geodesic extended mode (GEM), due to its connection with geodesic acoustic mode physics (see Section~\ref{sec:longtail_discussion}).

\begin{figure}
    \centering
    \begin{subfigure}[t]{0.49\columnwidth}
            \centering
            \includegraphics[width=\textwidth, trim={0.6cm 0.8cm 0.4cm 0.0cm}, clip]{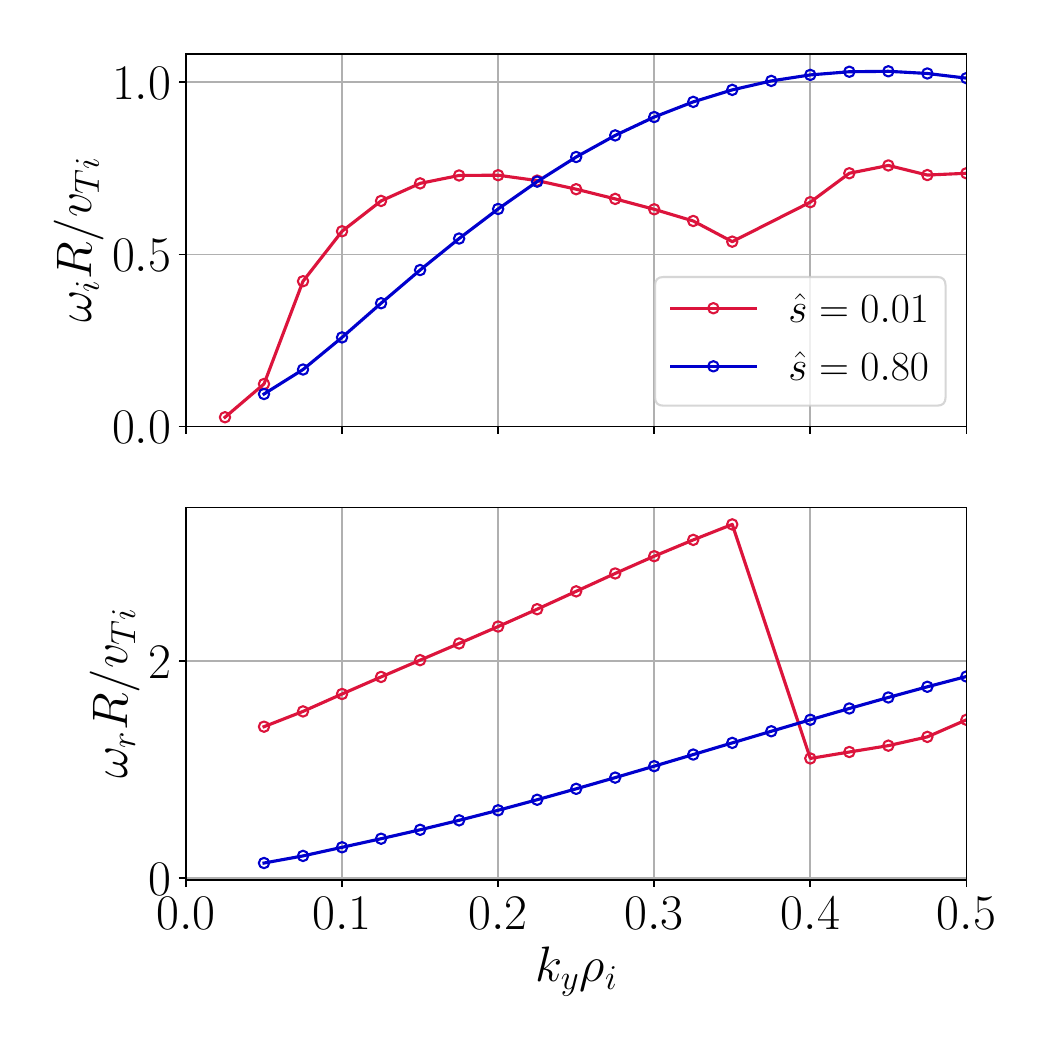}
        \caption{Growth rate (top) and real frequency (bottom) as a function of the binormal wavenumber $k_y \rho_i$.}
        \label{fig:tailmode_omega_shat_small_large}
    \end{subfigure}
   \begin{subfigure}[t]{0.49\columnwidth}
            \centering
            \includegraphics[width=\textwidth, trim={0.6cm 0.8cm 0.4cm 0.0cm}, clip]{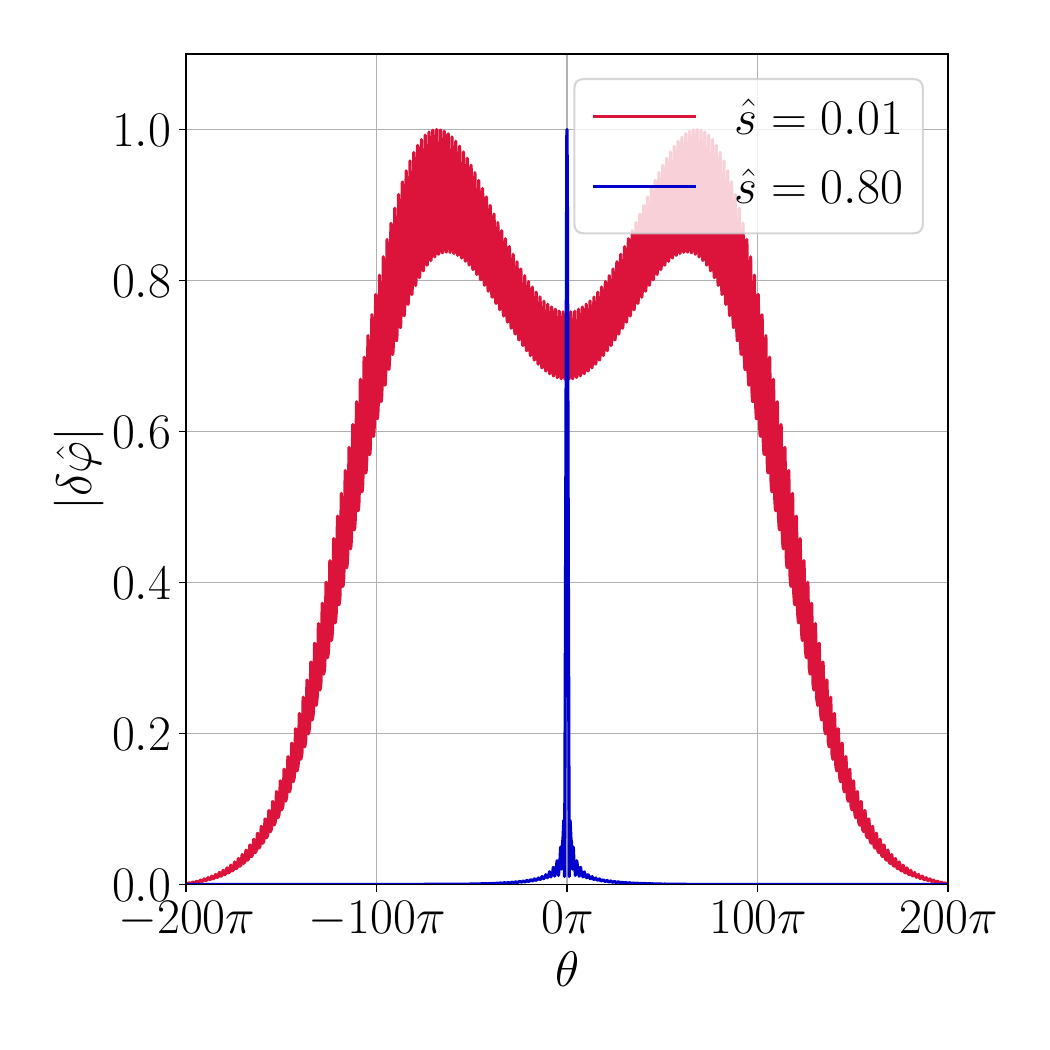}
        \caption{Potential fluctuation ballooning structure along the magnetic field for $k_y \rho_i = 0.1$.}
        \label{fig:tailmode_phi_shat_small_large}
    \end{subfigure}
    
    \caption[Ion-scale modes at small and large magnetic shear]{Complex frequency (a) and eigenmode structure for $k_y \rho_i = 0.1$ (b) at small and large magnetic shear, $\hat s =0.01$ and $\hat s = 0.8$. The data are obtained from gyrokinetic simulations using the \texttt{stella} code \citep{barnes_stella_2019} for a tokamak with circular cross-section. The safety factor is $q=2.38$, the electron-to-ion mass ratio is $m_e/m_i = 1/1836$, and a pure ion temperature gradient drive is considered, $a/L_{Ti}=4$ and ${a/L_{Te}=a/L_n = 0}$, where $a$ is the plasma minor radius. All other parameters are identical to those presented in Section~\ref{sec:longtail_sims_numerical}.}
    \label{fig:tailmode_shat_small_large}
\end{figure}

In Figure~\ref{fig:tailmode_shat_small_large}, we show representative gyrokinetic simulations exhibiting GEMs. The GEMs are indeed very extended compared with modes of a larger magnetic shear case, see Figure~\ref{fig:tailmode_phi_shat_small_large}. Furthermore, compared to the localised modes found at larger magnetic shear $\hat s \approx 0.8$, the GEMs oscillate rapidly in time and have a larger growth rate at small binormal wavenumbers.

The effect of low magnetic shear values on non-adiabatic passing electron physics has been studied in a series of recent papers by \cite{volcokas_ultra_2023, volcokas_numerical_2024, volcokas_numerical_2024-1, volcokas_turbulence-generated_2025}. Small values of magnetic shear are a common occurrence in stellarator magnetic fields, for example in stellarators optimised for quasisymmetry \citep{landreman_magnetic_2022, nies_exploration_2024, buller_family_2025}. In tokamaks, small magnetic shear values are especially relevant for reverse-shear operating scenarios, where the safety factor has a minimum away from the magnetic axis (in common tokamak operating scenarios, the safety factor monotonically increases from the core to the edge). In some reverse-shear discharges, large internal transport barriers were observed \citep[see e.g.][]{levinton_improved_1995, strait_enhanced_1995, eriksson_discharges_2002}, with greatly improved plasma confinement. It has been suggested by \cite{volcokas_turbulence-generated_2025} that these transport barriers arise due to the flattening of the safety factor profile caused by the self-interaction of extended modes, further motivating the study of the GEMs. We note that extended modes (though typically less extended than the modes presented in this paper) were also previously observed in linear gyrokinetic simulations at low magnetic shear for electrostatic instabilities in tokamaks and stellarators \citep{zocco_threshold_2018}, and for electromagnetic instabilities in tokamaks \citep{mulholland_resonant_2025}.

This paper is structured as follows. We first present the theory of extended modes at small magnetic shear in Section~\ref{sec:longtail_theory}. The theory is then validated against gyrokinetic simulations in Section~\ref{sec:longtail_sims}, which will also reveal the parametric dependences of the GEMs. We then elucidate the physics of the GEMs in Section~\ref{sec:longtail_discussion}. In Section~\ref{sec:longtail_precise_QA}, we show that GEMs are important to understand linear and nonlinear gyrokinetic simulations in stellarators by considering the precise quasi-axisymmetric stellarator of \cite{landreman_magnetic_2022}. Finally, we conclude and discuss our results in Section~\ref{sec:longtail_conclusion}.

\section{Theory of extended modes}\label{sec:longtail_theory}

In Section~\ref{sec:linear_GK}, we review the theory of gyrokinetics used to describe microinstabilities in tokamak and stellarator plasmas. We introduce the multiscale expansion of the gyrokinetic equation along the field line in Section~\ref{sec:longtail_theory_orderings}. We derive the ion and electron contributions to quasineutrality in a particular convenient limit in Sections~\ref{sec:longtail_theory_ions} and \ref{sec:longtail_theory_propagator}, respectively. These allow us to derive the dispersion relation for extended tail modes in Section~\ref{sec:longtail_theory_dispersion}. In Section~\ref{sec:longtail_theory_weak_limit}, we consider a subsidiary limit of slow electron propagation over the scale of the mode and finally, in Section~\ref{sec:longtail_model_disp}, we consider a simplified dispersion relation assuming fluid ions and negligible electron propagation.

\subsection{Linear gyrokinetics in tokamaks and stellarators}\label{sec:linear_GK}

We consider a strongly magnetised plasma, i.e. the gyroradius $\rho_{s} = v_{T{s}}/\Omega_{s}$ is much smaller than the system size ($\rho_{s} \ll R$), with $\Omega_{s} = Z_{s} e B/m_{s}$ the Larmor frequency, $Z_{s} e$ the particle charge, $e$ the proton charge, and $B$ the magnetic field strength. The fast gyromotion of particles may then be averaged over to derive gyrokinetics \citep{catto_linearized_1978, frieman_nonlinear_1982}, a kinetic theory describing the motion of rings of charge. We consider $\delta f$-gyrokinetics: for each species, the particle distribution function $f_{s}$ may be decomposed as $f_{s} = F_{M{s}} + \delta f_{s}$, where the fluctuating distribution function ${\delta f_{s} \sim (\rho_{s}/R) F_{M{s}}}$ is small compared to the background Maxwellian
\begin{equation}
    F_{M{s}} = \frac{n_{s}}{\pi^{3/2} v_{T{s}}^3} \exp\left( - \frac{v^2}{v_{T{s}}^2} \right),
\end{equation}
with $n_{s}$ the background density and $v$ the particle speed. The spatial derivatives of the fluctuating distribution function are ordered as
\begin{equation}
    \abs{\bhat \cdot \nabla \ln \delta f_{s}} \sim 1/R, \qquad \abs{\bhat \times \nabla \ln \delta f_{s}} \sim 1/\rho_{s},
\end{equation}
corresponding to anisotropic turbulent eddies that are elongated along the magnetic field but are narrow in the direction perpendicular to it. Here, $\bhat = \bs{B}/B$ is the unit magnetic field vector. Finally, the derivatives in velocity space and with respect to time are ordered as
\begin{equation}
     \abs{\partial_{\bs{v}} \ln \delta f_{s}} \sim 1/v_{T{s}}, \qquad \abs{\partial_t \ln \delta f_{s}} \sim v_{T{s}}/R.
\end{equation}

It is convenient to express the gyrokinetic equation in terms of the non-adiabatic contribution to the distribution function $h_{s}$, which satisfies
\begin{equation}
    \delta f_{s}(\bs{r}, \sigma, \epsGK, \muGK, \vartheta,t) = h_{s}(\bs{R},  \sigma, \epsGK, \muGK, t) - \frac{Z_{s} e \delta\varphi(\bs{r})}{T_{s}} F_{M{s}},
\end{equation}
with $\delta\varphi$ the fluctuating electrostatic potential. Here, we have defined the sign of the parallel velocity $\sigma=v_\parallel / \abs{v_\parallel}$ with $v_\parallel=\bs{v}\cdot\bhat$, the particle energy ${\epsGK=m_{s} v^2/2}$, the magnetic moment $\muGK = m_{s} v_\perp^2/(2B)$, the gyrophase $\vartheta$ measuring the angle between $\bs{v}_\perp = \bs{v}-v_\parallel\bhat$ and an arbitrarily defined coordinate axis perpendicular to the magnetic field, and finally the guiding centre position
\begin{equation}
    \bs{R} = \bs{r} - \frac{\bhat\times\bs{v}}{\Omega_{s}},
\end{equation}
which is shifted from the real space position $\bs{r}$ by the gyroradius vector. We note that the non-adiabatic piece $h_s$ does not depend on the gyrophase $\vartheta$.

We consider the linearised, electrostatic, collisionless, and low-flow limit of the gyrokinetic equation, given by
\begin{equation} \label{eq:GK}
    \frac{\partial h_{{s}}}{\partial t} + \left(v_\parallel \bhat + \vmtildea\right) \cdot \frac{\partial h_{{s}}}{\partial \RGK} + \langle \bs{v}_E \rangle_{\RGK}\cdot \frac{\partial  F_{M{s}}}{\partial \RGK} = \frac{Z_{s} e}{T_{s}}   \dd{\gyroavgR{\delta \varphi}}{t} F_{M{s}},
\end{equation}
where $\gyroavgR{}$ is a gyro-average (average over the gyrophase $\vartheta$) at fixed guiding centre position $\bs{R}$. Here, we defined the fluctuating $\bs{E}\times\bs{B}$ drift velocity
\begin{equation}
    \bs{v}_E = \frac{\bhat\times\nabla\delta\varphi}{B},
\end{equation}
and the magnetic drift velocity
\begin{equation}
    \vmtildea  = \frac{\bhat}{\Omega_{s}} \times \left( v_\parallel^2 \bhat\cdot\nabla\bhat + \frac{v_\perp^2}{2}\nabla\ln B \right),
\end{equation}
which is the combination of the curvature and $\nabla B$ drifts.

The fluctuating electrostatic potential $\delta\varphi$ entering the gyrokinetic equation \eqref{eq:GK} is obtained self-consistently from the quasineutrality equation
\begin{equation} \label{eq:QN}
    0 = \sum_{s} Z_{s} \int\mathrm{d}^3 v\, \delta f_{s} = - \sum_{s} \frac{Z_{s}^2 e n_{s}}{T_{s}} \delta\varphi + \sum_{s} Z_{s} \int\mathrm{d}^3 v\, \langle h_{{s}}\rangle_{\bs{r}},
\end{equation}
where $\langle \rangle_{\bs{r}}$ denotes a gyro-average at fixed real space position $\bs{r}$.

In a tokamak or stellarator, it is convenient to use the coordinate system $(\psi, \alphafl, \theta)$, where $\psi$ is the poloidal flux labeling flux surfaces and $\alphafl = \zeta - q\theta + \nu(\theta,\zeta)$ is a field line label defined such that $\bs{B}=\nabla\psi\times\nabla\alphafl$. Here, $\nu$ is a single-valued function of the toroidal and poloidal angles $\zeta$ and $\theta$, and $q=q(\psi)$ is the safety factor measuring the average pitch of magnetic field lines on a flux surface.

We consider a narrow flux tube centered on the flux surface $\psi_0$ and field line $\alphaflz$. The flux tube domain is parameterised in the perpendicular directions by
\begin{equation}
    x = (\psi-\psi_0) \DD{x}{\psi}, \qquad y = (\alphafl-\alphaflz) \DD{y}{\alphafl},
\end{equation}
with the radial coordinate $x$ labeling flux surfaces, the binormal coordinate $y$ labeling field lines on a flux surface. The constants $\mathrm{d} x/\mathrm{d}\psi$ and $\mathrm{d}y/\mathrm{d}\alphafl$ are arbitrary and are chosen differently in different formulations and codes.

Because the flux tube is narrow in the direction perpendicular to $\bs{B}$ compared to the system scale length, statistical periodicity of the fluctuations may be invoked to justify the use of periodic boundary conditions in the directions perpendicular to $\bs{B}$. We may therefore consider the fluctuations in Fourier space, parameterised by a binormal wavenumber $k_y$ and radial wavenumber $k_x$. We also consider Fourier eigenmodes in time, allowing us to express the distribution function and fluctuating electrostatic potential as
\begin{align}
    h_s(t, X, Y, \theta, \varepsilon, \mu, \sigma) & = \sum_{k_x, k_y, \omega} \hat h_s(\omega, k_x, k_y, \theta, \varepsilon, \mu, \sigma)e^{-i\omega t+ik_x X+ik_y Y}, \\
        \delta\varphi(t, x, y, \theta) & = \sum_{k_x, k_y, \omega}  \delta\hat\varphi(\omega, \theta, k_x, k_y)e^{-i\omega t+ik_x x+ik_y y},
\end{align}
where $X$ and $Y$ are the radial and binormal coordinates of the guiding centre.

It will prove convenient to express the radial wavenumber through the quantity
\begin{equation} \label{eq:theta0}
    \theta_0 = \frac{k_x}{k_y} \left(\DD{q}{x} \DD{y}{\alphafl} \right)^{-1}.
\end{equation}
The perpendicular wavevector may then be written as
\begin{align} \label{eq:kperp_tail_general}
    \bs{k}_\perp & = k_x \nabla x + k_y \nabla y \nonumber\\
    & = k_y \DD{y}{\alphafl} \left( \nabla\zeta\left(1 + \partial_\zeta\nu\right) + \nabla\theta \left( - q + \partial_\theta\nu \right ) + \nabla x \DD{q}{x} \left( \theta_0-\theta \right) \right),
\end{align}
where the last term captures the secular increase in $\theta$ due to magnetic shear 
\begin{equation}
    \hat s = \frac{\mathrm{d}\ln q}{\mathrm{d}\ln \psi} = \frac{\psi_0}{q(\psi_0)} \frac{\mathrm{d}x}{\mathrm{d}\psi} \DD{q}{x}.
\end{equation}

In Fourier space, the linearised GK equation \eqref{eq:GK} is given by
\begin{equation} \label{eq:GK_eqn_ballooning}
    (\omega - \tilde{\omega}_{M{s}}) \hat h_{s}  + i v_\parallel \bhat\cdot\nabla\theta \dd{\hat h_{s}}{\theta} = (\omega - \tilde\omega_{*{s}}) J_{0{s}} F_{M{s}} \frac{Z_{s} e \delta\hat\varphi}{T_{s}},
\end{equation}
and the quasineutrality equation \eqref{eq:QN} becomes
\begin{equation} \label{eq:QN_ballooning}
    0 = \sum_{s} Z_{s} \delta\hat n_{s} = - \sum_{s} \frac{Z_{s}^2 e n_{s}}{T_{s}} \delta\hat\varphi + \sum_{s} Z_{s} \int\mathrm{d}^3 v\, J_{0{s}}\hat h_{{s}}.
\end{equation}
Equation \eqref{eq:GK_eqn_ballooning} is an ordinary differential equation in $\theta$, with $\theta \in (-\infty, \infty)$. For passing particles, we assume that there is no inflow of particles from infinity, i.e.
\begin{equation} \label{eq:BC_no_inflow_infty}
    \lim_{\theta \rightarrow -\sigma \infty}\hat h_{s}^\mathrm{p}(\theta,\sigma) = 0.
\end{equation}
For trapped particles, the distribution function must be independent of $\sigma$ at the bounce points $\theta_b^{\pm}$, i.e.
\begin{equation}
    \hat h_{s}^\mathrm{t}(\theta_b^\pm,\sigma=1) = \hat h_{s}^\mathrm{t}(\theta_b^\pm,\sigma=-1).
\end{equation}
In equation \eqref{eq:GK_eqn_ballooning}, the magnetic drift frequency is defined as
\begin{equation} \label{eq:omega_Malpha}
    \tilde{\omega}_{M{s}} = \bs{k}_\perp \cdot \tilde{\bs{v}}_{M{s}},
\end{equation}
and the diamagnetic frequency is given by
\begin{align}\label{eq:omega_staralpha}
    \tilde\omega_{*{s}} & = \omega_{*{s}} + \omega_{*{s}}^T \left( \frac{v^2}{v_{T{s}}^2} - \frac{3}{2} \right) \nonumber\\
    & = -k_y \DD{y}{\alphafl} \frac{T_{s}}{Z_{s} e}   \left( \DD{\ln n_{s}}{\psi}  + \DD{\ln T_{s}}{ \psi} \left( \frac{v^2}{v_{T{s}}^2} - \frac{3}{2} \right) \right),
\end{align}
which implicitly defines $\omega_{*{s}}$ and $\omega_{*{s}}^T$. Finite Larmor radius effects are captured by the Bessel functions of the first kind
\begin{equation}
    J_{0{s}} = J_0 \left( \frac{k_\perp v_\perp}{\Omega_{s}} \right).
\end{equation}

\subsection{Ordering and multiscale expansion}\label{sec:longtail_theory_orderings}

As discussed previously, the modes we seek to describe become very extended along the magnetic field as the magnetic shear is reduced. However, these modes will also have some variation over smaller scales ($\theta \sim 1$) due to the variation of the magnetic geometry, which enters $\tilde\omega_{Ms}$ and $J_{0s}$. To capture both of these behaviours, we follow in the footsteps of \cite{hardman_extended_2022} and consider a multiscale expansion in $\theta$ of the distribution function, $\hat h_{s}(\theta) \rightarrow \hat h_{s}(\theta_s, \theta_f)$. Here, $\theta_f$ captures the fast variation of geometric quantities, typically over a single poloidal turn, while $\theta_s$ captures the slower variation due to the magnetic shear.

We use the small electron-to-ion mass ratio to define the expansion parameter 
\begin{equation}
        \epsilon = \sqrt{m_e/m_i} \ll 1
\end{equation}
and we order the magnetic shear to be small in $\epsilon$, i.e.
\begin{equation}
    \hat s \sim \epsilon.
\end{equation}
To leading order in $\epsilon$, we may approximate the perpendicular wavenumber vector \eqref{eq:kperp_tail_general} as
\begin{equation} \label{eq:kperp_tail}
    \bs{k}_\perp = k_y \DD{y}{\alphafl} \left( \nabla\zeta\left(1 + \partial_\zeta\nu\right) + \nabla\theta \left( - q + \partial_\theta\nu \right ) - \nabla x \DD{q}{x} \theta_s \right),
\end{equation}
where the secular variation due to the magnetic shear is captured through $\theta_s$ only, and $\nabla \zeta $, $\nabla \theta$, $\partial_\zeta \nu$, $\partial_\theta \nu$, and $\nabla x$ are functions of $\theta_f$ only. We also neglected the contribution from $\theta_0$ to the secular term in \eqref{eq:kperp_tail_general}, i.e. we assumed $\theta_s \gg \theta_0 \sim 1$. This may be justified rigorously in a tokamak, where $\theta_0 \in [0,2\pi)$ without loss of generality.

The terms in the gyrokinetic equation \eqref{eq:GK_eqn_ballooning} for ions and electrons are ordered as
\begin{equation} \label{eq:longtail_multiscale_ordering}
    \epsilon\, v_{Te} \bhat \cdot\nabla\theta \dd{\ln \hat h_e}{\theta_f} \sim v_{Te} \bhat \cdot\nabla\theta \dd{\ln \hat h_e}{\theta_s} \sim \tilde\omega_{*e} \sim \tilde\omega_{Me} \sim v_{Ti} \bhat \cdot\nabla\theta \dd{\ln \hat h_i}{\theta_f} \sim \tilde\omega_{*i} \sim \tilde \omega_{Mi} \sim \omega.
\end{equation}
The electrons propagate quickly over a distance $\theta_f$, while their propagation speed over a distance $\theta_s$ is maximally ordered with the mode frequency, the magnetic drift frequencies, the diamagnetic frequencies, and the ion propagation speed over $\theta_f$.

In what follows, we consider a single ion species for ease of notation, though the theory may easily be generalised to multiple species. We study ion-scale modes with $k_\perp \rho_e \ll k_\perp \rho_i \sim 1$. We therefore neglect electron FLR effects, such that $J_{0e}=1$.

\subsection{Local ion response}\label{sec:longtail_theory_ions}

With the ordering \eqref{eq:longtail_multiscale_ordering}, the gyrokinetic equation \eqref{eq:GK_eqn_ballooning} for the ions is given by
\begin{equation} \label{eq:GK_eqn_ballooning_ions}
        (\omega - \tilde{\omega}_{Mi}) \hat h_i  + i v_\parallel \bhat\cdot\nabla\theta \dd{\hat h_i}{\theta_f} = (\omega - \tilde\omega_{*i}) J_{0i} F_{Mi} \frac{Z_i e \delta\hat\varphi}{T_i}.
\end{equation}

To make further progress, we neglect the ion parallel streaming term. This approximation is often appropriate for the geodesic extended modes, which oscillate rapidly with frequency $\omega \sim v_{Ti}/R$, even for small $k_y \rho_i$ (see Figure~\ref{fig:tailmode_omega_shat_small_large}). The ion parallel streaming may be neglected when the connection length is sufficiently long; we discuss this approximation in Section~\ref{sec:longtail_limits_validity}. Neglecting the ion parallel streaming, the ion gyrokinetic equation \eqref{eq:GK_eqn_ballooning_ions} simplifies to an algebraic equation, which may be rearranged to obtain
\begin{equation}
     \hat h_i  = \frac{\omega - \tilde\omega_{*i}}{\omega - \tilde{\omega}_{Mi}} J_{0i} F_{Mi} \frac{Z_i e \delta\hat\varphi}{T_i}.
\end{equation}
The ion density fluctuations may therefore be written as
\begin{equation} \label{eq:longtail_ni}
    \frac{\delta \hat n_i}{n_i} = -\frac{Z_i e \delta\hat\varphi}{T_i} + \frac{1}{n_i}\int\mathrm{d}^3 v\, J_{0i}\hat h_{i} = -\frac{Z_i e \delta\hat\varphi }{T_i} (1-\mathcal{M}_i),
\end{equation}
where we defined the ion response function
\begin{equation}\label{eq:longtail_Mi}
    \mathcal{M}_i = \frac{1}{n_i}\int\mathrm{d}^3 v\, \frac{\omega - \tilde\omega_{*i}}{\omega - \tilde{\omega}_{Mi}} J_{0i}^2 F_{Mi},
\end{equation}
which depends on both $\theta_f$ and $\theta_s$ through the Bessel functions and the magnetic drift frequency.

To numerically evaluate $\mathcal{M}_i$, it is convenient to express the velocity integral as a one-dimensional integral \citep{terry_kinetic_1982}. We consider unstable modes (Im$(\omega)>0$), such that the resonant denominator in \eqref{eq:longtail_Mi} may be written as
\begin{equation}
    \frac{1}{ \omega - \tilde{\omega}_{Mi}} = -i \int_0^\infty \mathrm{d}\Lambda\, \exp\left[ i \Lambda(\omega-\tilde{\omega}_{Mi}) \right].
\end{equation}
The ion response function \eqref{eq:longtail_Mi} may then be reduced to (see also \cite{biglari_toroidal_1989, zocco_threshold_2018, parisi_toroidal_2020})
\begin{align}
    \mathcal{M}_i & = -i \int_0^\infty\mathrm{d}\Lambda   \frac{e^{i\Lambda \omega}}{\sqrt{1+i\Lambda \omegakappai}} \frac{1}{1+i \Lambda \omeganablaBi /2} \Bigg\{ \tilde b_i \left( \Gamma_0(\tilde b_i) - \Gamma_1(\tilde b_i) \right) \frac{\omega_{*i}^T}{1 + i \Lambda \omeganablaBi/2} \nonumber \\ 
    & + \Gamma_0(\tilde b_i) \left( \omega - \omega_{*i} - \omega_{*i}^T \left( \frac{1/2}{1+i \Lambda \omegakappai} + \frac{1}{1+i\Lambda \omeganablaBi/2} - \frac{3}{2} \right) \right) \Bigg\}, \label{eq:curlyM}
\end{align}
where we have defined the curvature drift frequency
\begin{equation}
    \omegakappai =  \frac{v_{Ti}^2}{\Omega_i} \bs{k}_\perp \cdot  \bhat \times(\bhat\cdot\nabla \bhat),
\end{equation}
the $\nabla B$ drift frequency
\begin{equation}
    \omeganablaBi = \frac{v_{Ti}^2}{\Omega_i}\bs{k}_\perp \cdot  \bhat \times \nabla\ln B,
\end{equation}
and the quantity
\begin{equation}
    \tilde b_i = \frac{k_\perp^2 \rho_i^2}{2} \frac{1}{1 + i \Lambda \omeganablaBi/2},
\end{equation}
which sets the strength of ion FLR effects. These enter in \eqref{eq:curlyM} through the functions $\Gamma_n(\tilde b_i) = I_n(\tilde b_i) e^{-\tilde b_i}$, where $I_n$ is the modified Bessel function of the first kind.

The integral over $\Lambda$ in \eqref{eq:curlyM} is well-behaved as long as Re$(\Lambda) \rightarrow +\infty$ and Im$(\omega) > 0$, such that the complex exponential $e^{i\Lambda \omega}$ decays. Numerically, we find it sufficient to carry out the integration along the positive real axis of $\Lambda$, though other integration paths may also be considered \citep[see e.g.][]{parisi_toroidal_2020}.

\subsection{Electron propagator}\label{sec:longtail_theory_propagator}

The electron distribution function is expanded in the parameter $\epsilon$ as
\begin{equation}
    \hat h_e = \hat h_{e,(0)} + \hat h_{e,(1)} + ...,
\end{equation}
where $\hat h_{e,(n)} \sim \epsilon^n \hat h_e$. With the ordering \eqref{eq:longtail_multiscale_ordering}, the electron gyrokinetic equation is given to leading order by
\begin{equation} \label{eq:tail_GK_e_zeroth}
    i v_\parallel \bhat\cdot\nabla\theta \dd{\hat h_{e,(0)}}{\theta_f} = 0,
\end{equation}
and to next order by
\begin{equation}\label{eq:tail_GK_e_first}
    (\omega - \tilde{\omega}_{M e}) \hat h_{e,(0)}  + i v_\parallel \bhat\cdot\nabla\theta \dd{\hat h_{e,(0)}}{\theta_s} + i v_\parallel \bhat\cdot\nabla\theta \dd{\hat h_{e,(1)}}{\theta_f} = -(\omega - \tilde\omega_{*e}) F_{Me} \frac{e \delta\hat\varphi}{T_e}.
\end{equation}

We now define the transit average over the fast coordinate $\theta_f$, 
\begin{equation}\label{eq:longtail_transit_avg}
    \langle A \rangle_{\tau_f} =
    \begin{cases}
        \left( \int_{\theta_f^-}^{\theta_f^+} \mathrm{d}\theta_f \frac{A}{v_\smallparallel \bhat\cdot\nabla\theta}  \right) \Big/ \left( \int_{\theta_f^-}^{\theta_f^+} \mathrm{d}\theta_f \frac{1}{v_\smallparallel \bhat\cdot\nabla\theta}  \right)  \qquad & (\text{passing}),\\
        \left(\sum_\sigma \int_{\theta_b^-}^{\theta_b^+} \mathrm{d}\theta_f \frac{A}{\abs{v_\smallparallel} \bhat\cdot\nabla\theta}  \right) \Big/ \left(2 \int_{\theta_b^-}^{\theta_b^+} \mathrm{d}\theta_f \frac{1}{\abs{v_\smallparallel} \bhat\cdot\nabla\theta}  \right)   \qquad & (\text{trapped}).
    \end{cases}
\end{equation}
Here, $\theta_b^\pm$ are the bounce points of trapped particles. We proceed to discuss the limits of integration $\theta_f^\pm$ for the passing particle transit average over the fast coordinate. We assume the existence of a periodicity length in the fast variable $\theta_f$ given by $\theta_f^+ - \theta_f^-$. Thus, the passing electron distribution function is assumed to satisfy
\begin{equation} \label{eq:tail_he_periodicity_thetaf}
    \hat h_e^\mathrm{p}(\theta_f = \theta_f^-) = \hat h_e^\mathrm{p}(\theta_f =\theta_f^+).
\end{equation}
This assumption is justified in tokamaks, where $\theta_f^\pm = \pm \pi$ as all geometric quantities are periodic in $\theta$, and on rational flux surfaces in stellarators, where $\theta_f^\pm = \pm N\pi$, as geometric quantities are periodic after $N$ poloidal turns if $q =  M/N$ with $M,N\in \mathbb{N}$. On an irrational flux surface in a stellarator, finding $\theta_f^\pm$ such that \eqref{eq:tail_he_periodicity_thetaf} is satisfied requires the flux tube to have sufficiently sampled the geometry of the flux surface.

From equation \eqref{eq:tail_GK_e_zeroth}, the leading order electron distribution function satisfies
\begin{equation}
    \hat h_{e,(0)} = \left\langle \hat h_{e,(0)} \right\rangle_{\tau_f}
\end{equation}
and is determined as a solvability condition of \eqref{eq:tail_GK_e_first}, whose transit average over the fast coordinate is given by
\begin{equation} \label{eq:tail_GK_e_first_transit_avg}
    (\omega - \langle \tilde{\omega}_{M e} \rangle_{\tau_f}) \hat h_{e,(0)}  + i \langle v_\parallel \bhat\cdot\nabla\theta\rangle_{\tau_f} \dd{\hat h_{e,(0)}}{\theta_s} = -(\omega - \tilde\omega_{*e}) F_{Me} \frac{e \langle \delta\hat\varphi \rangle_{\tau_f}}{T_e}.
\end{equation}

As trapped particles are localised in magnetic field wells, they cannot sample the slow variation in $\theta_s$. Indeed, $\langle v_\parallel \bhat\cdot\nabla\theta \rangle_{\tau_f} = 0$ for trapped particles, and the trapped electron distribution $\hat h_{e,(0)}^\mathrm{t}$ is therefore given by
\begin{equation}
    \hat h_{e,(0)}^\mathrm{t} = -\frac{\omega - \tilde\omega_{*e}}{\omega - \langle \tilde{\omega}_{M e} \rangle_{\tau_f}} F_{Me} \frac{e \langle \delta\hat\varphi \rangle_{\tau_f}}{T_e}.
\end{equation}

To determine the passing electron distribution function $h_{e,(0)}^\mathrm{p}$, we first define
\begin{equation} \label{eq:longtail_phie}
    \phi_e(\theta_s) = \frac{\omega- \langle \tilde{\omega}_{M e} \rangle_{\tau_f}}{\langle v_\parallel \bhat\cdot\nabla\theta\rangle_{\tau_f}}
\end{equation}
and
\begin{equation}
    H_e = \hat h_{e,(0)}^\mathrm{p} \exp\left(-i \int^{\theta_s}\mathrm{d}\theta_s'\,\phi_e(\theta_s')\right).
\end{equation}
Equation \eqref{eq:tail_GK_e_first_transit_avg} for passing particles may then be written as
\begin{equation}
    i \langle v_\parallel \bhat\cdot\nabla\theta\rangle_{\tau_f} \dd{H_e}{\theta_s} = -(\omega - \tilde\omega_{*e}) F_{Me} \frac{e \langle \delta\hat\varphi \rangle_{\tau_f}}{T_e} \exp\left(-i \int^{\theta_s}\mathrm{d}\theta_s'\,\phi_e(\theta_s')\right).
\end{equation}
Integrating from infinity and using the boundary condition \eqref{eq:BC_no_inflow_infty}, the leading order electron distribution function is derived to be
\begin{equation} \label{eq:tail_he0P}
    \hat h_{e,(0)}^\mathrm{p}(\theta_s) = i \frac{\omega - \tilde\omega_{*e}}{\langle v_\parallel \bhat\cdot\nabla\theta\rangle_{\tau_f}} F_{Me} \int_{-\sigma \infty}^{\theta_s} \mathrm{d}\theta_s' \frac{e \langle \delta\hat\varphi \rangle_{\tau_f}(\theta_s')}{T_e} \exp\left(i \int^{\theta_s}_{\theta_s'}\mathrm{d}\theta_s''\,\phi_e(\theta_s'')\right).
\end{equation}

To make further progress, we now make two simplifying assumptions. First, we consider a large aspect ratio configuration, i.e. the variation of $B$ and $v_\parallel$ in $\theta$ are assumed to be negligible. Therefore, the number of trapped particles is negligibly small, and transit averages over the fast coordinate \eqref{eq:longtail_transit_avg} simplify to averages over $\theta_f$, 
\begin{equation} \label{eq:longtail_avg_thetaf}
    \langle A \rangle_{\tau_f} \rightarrow \langle A \rangle_{\theta_f} = \left( \int_{\theta_f^-}^{\theta_f^+} \mathrm{d}\theta_f \frac{A}{\bhat\cdot\nabla\theta} \right) \Bigg/ \left( \int_{\theta_f^-}^{\theta_f^+} \mathrm{d}\theta_f \frac{1}{\bhat\cdot\nabla\theta} \right).
\end{equation}

\sloppy Second, we assume that the transit-averaged magnetic drift frequency of electrons is negligible, $\langle \tilde \omega_{Me} \rangle_{\tau_f} \ll \omega$. This may be justified e.g. by considering small $k_y \rho_i \ll 1$ because $\tilde \omega_{Me} / \omega \sim k_y \rho_i$, where we have assumed $\omega \sim v_{Ti}/R$ --- we will show that this is the GEM frequency in Section~\ref{sec:longtail_sims}. Furthermore, in practice, the transit average (over the fast coordinate) of the magnetic drift is smaller than the local value of the magnetic drift. We discuss the validity of our assumptions in more detail in Section~\ref{sec:longtail_limits_validity}.

With the above assumptions, we only need to consider the passing electron distribution function \eqref{eq:tail_he0P}, which simplifies to
\begin{equation}
    \hat h_{e,(0)} = i \frac{\omega - \tilde\omega_{*e}}{v_\parallel \langle \bhat\cdot\nabla\theta\rangle_{\theta_f}} F_{Me} \int_{-\sigma \infty}^{\theta_s} \mathrm{d}\theta_s' \frac{e \langle \delta\hat\varphi \rangle_{\theta_f}(\theta_s')}{T_e} \exp\left(\frac{i \omega (\theta_s -\theta_s') }{v_\parallel \langle \bhat \cdot\nabla\theta \rangle_{\theta_f}}  \right).
\end{equation}

\begin{figure}
    \centering
    
    \begin{subfigure}[t]{0.49\linewidth}
            \centering
            \includegraphics[width=\textwidth, trim={1.6cm 1.5cm 0.6cm 2cm}, clip]{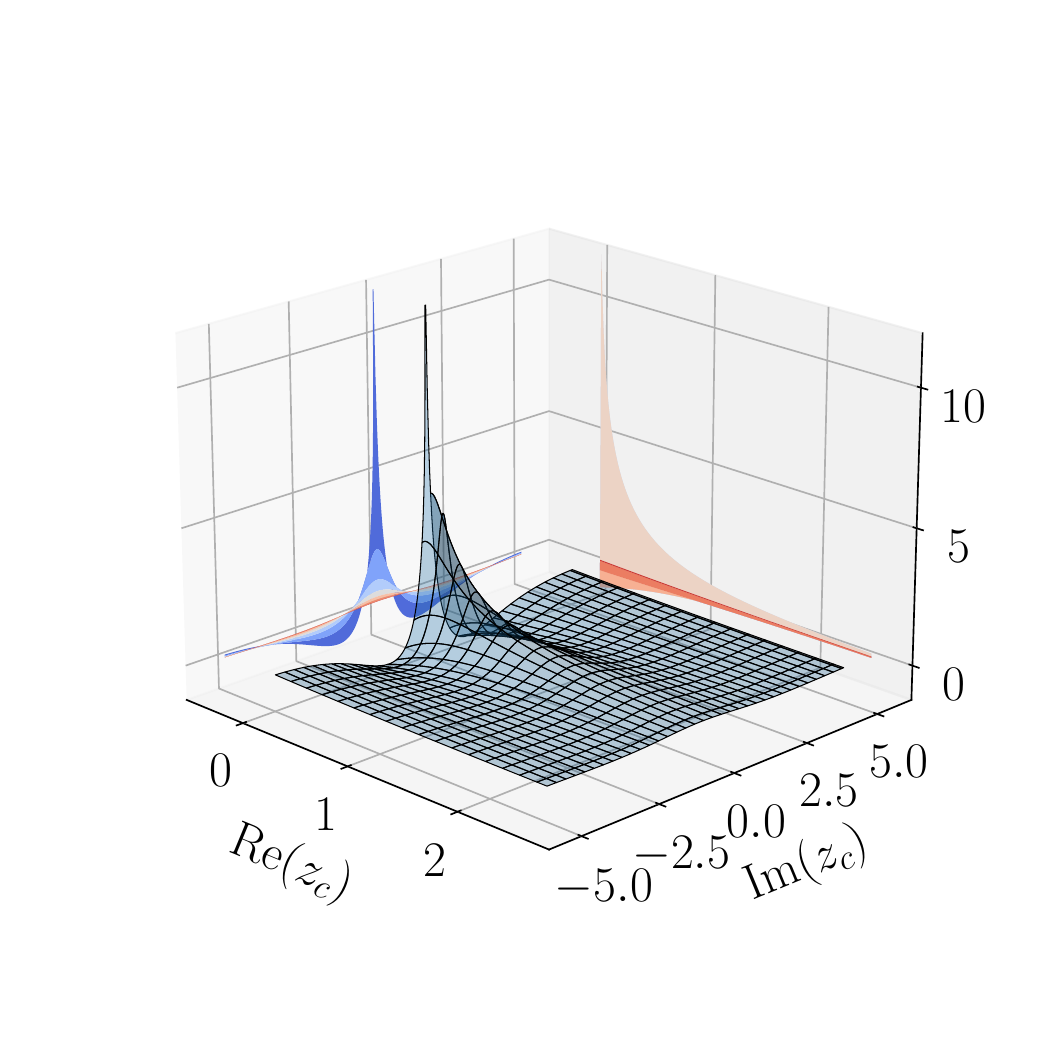}
        \caption{Re$\left(\MeijerG[\bigg]{0}{0}{0}{3}{\dots\\}{0,0,1/2}{\frac{z_c^2}{4}}\right)$}
    \end{subfigure}
    \begin{subfigure}[t]{0.49\linewidth}
            \centering
            \includegraphics[width=\textwidth, trim={1.6cm 1.5cm 0.6cm 2cm}, clip]{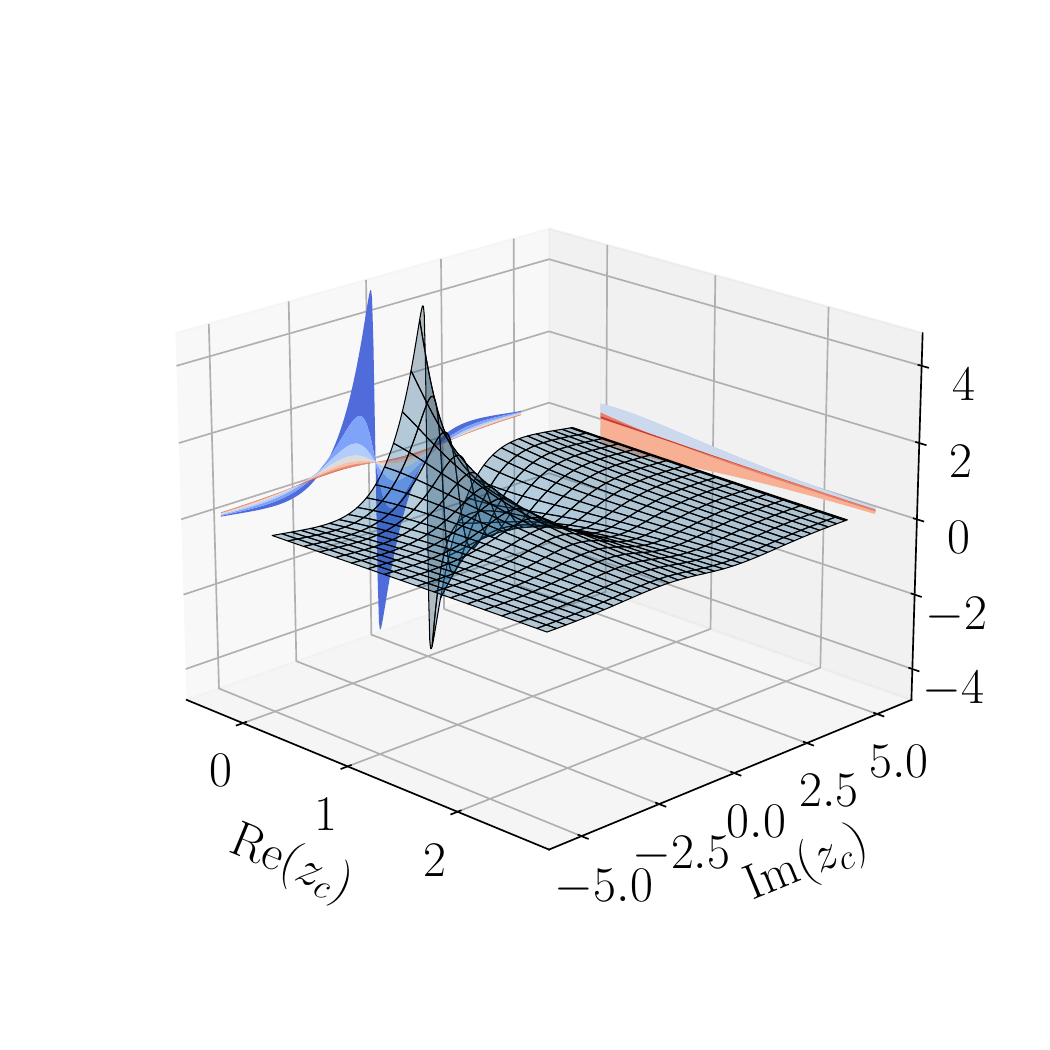}
        \caption{Im$\left(\MeijerG[\bigg]{0}{0}{0}{3}{\dots\\}{0,0,1/2}{\frac{z_c^2}{4}}\right)$}
    \end{subfigure}
    
    \begin{subfigure}[t]{0.49\linewidth}
            \centering
            \includegraphics[width=\textwidth, trim={1.6cm 1.5cm 0.6cm 2cm}, clip]{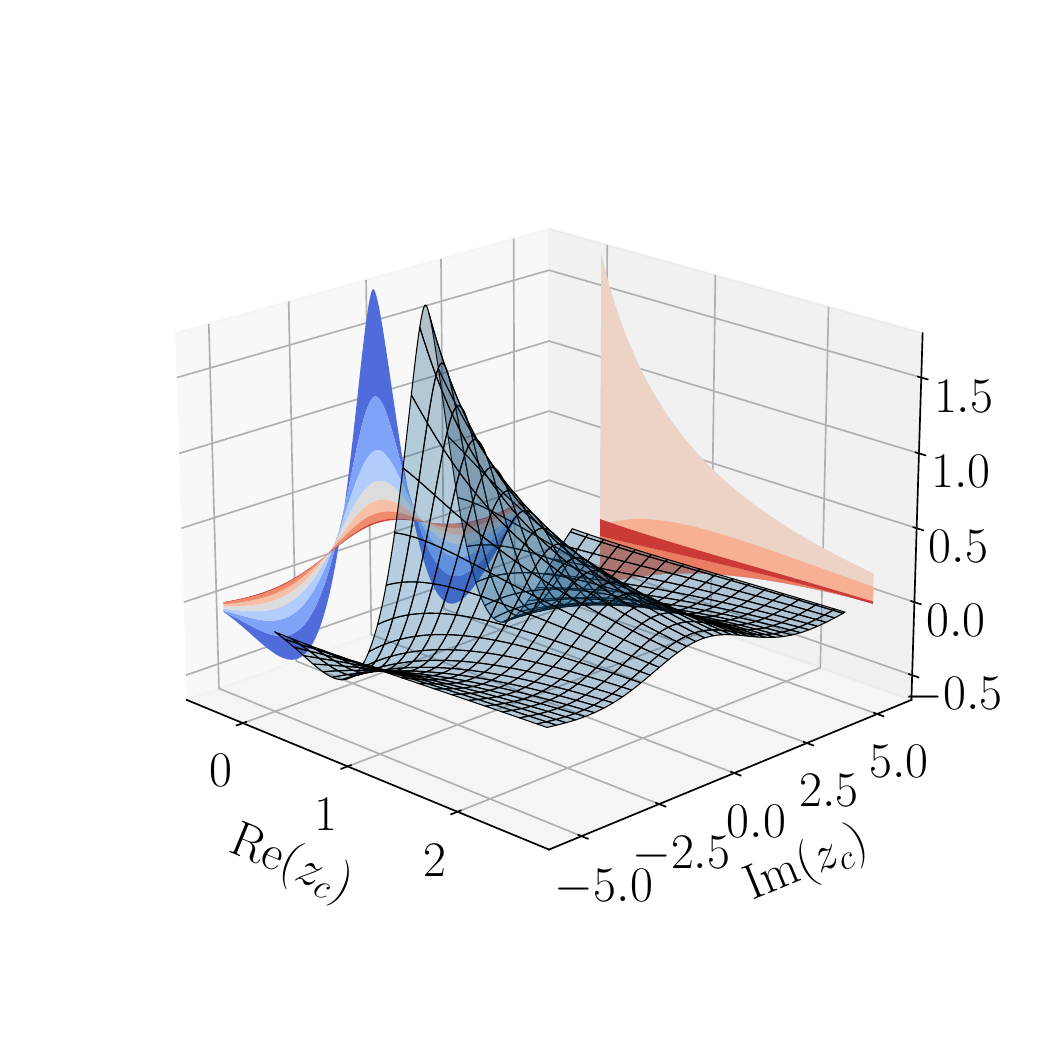}
        \caption{Re$\left(\MeijerG[\bigg]{0}{0}{0}{3}{\dots\\}{0,1/2,1}{\frac{z_c^2}{4}}\right)$}
    \end{subfigure}
    \begin{subfigure}[t]{0.49\linewidth}
            \centering
            \includegraphics[width=\textwidth, trim={1.6cm 1.5cm 0.6cm 2cm}, clip]{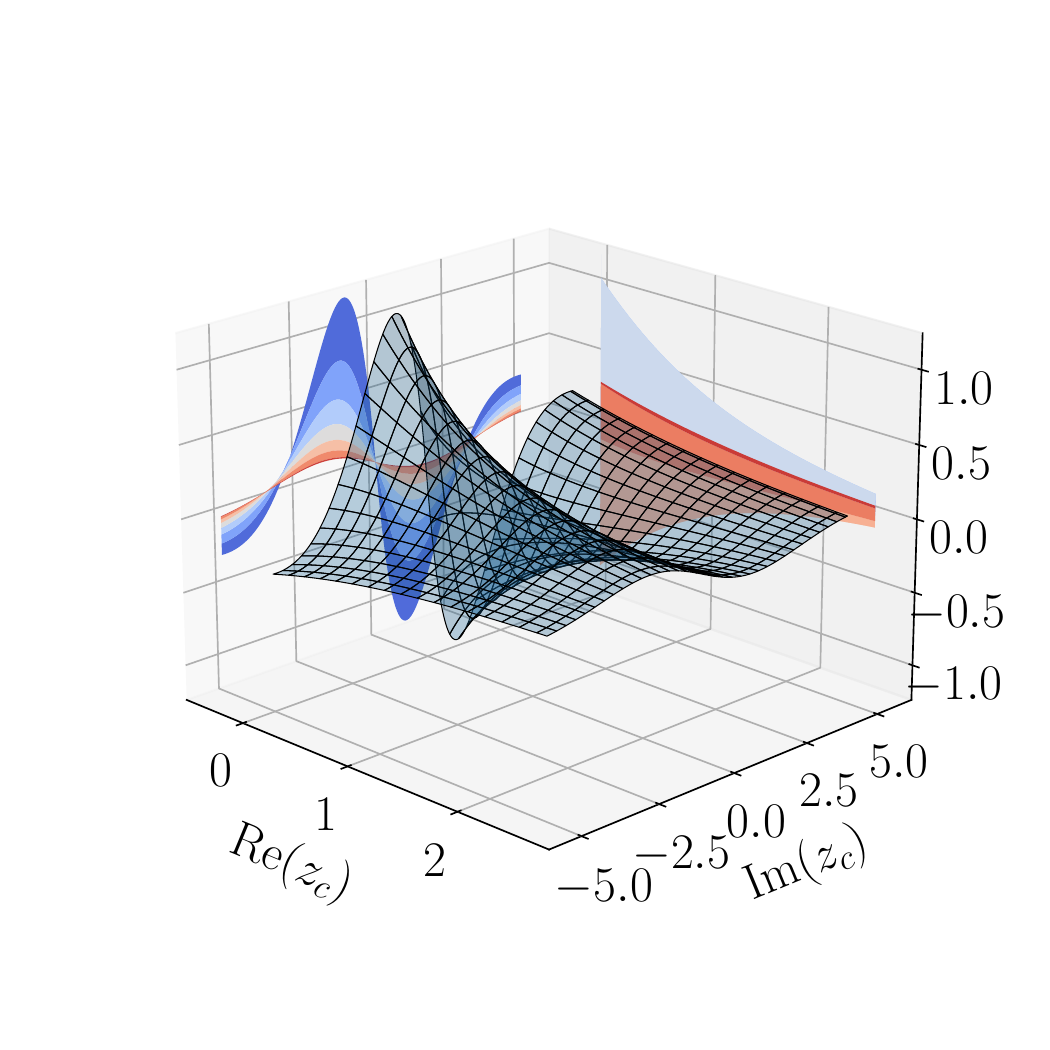}
        \caption{Im$\left(\MeijerG[\bigg]{0}{0}{0}{3}{\dots\\}{0,1/2,1}{\frac{z_c^2}{4}}\right)$}
    \end{subfigure}
    
    \caption[Visualisation of Meijer G-functions]{Meijer G-functions for the evaluation of the electron propagator \eqref{eq:Pe_MeijerG}. Only Re$(z_c)>0$ values are considered, corresponding to growing modes (Im $\omega=\omega_i>0$). One-dimensional cuts are projected on the sides as contours.}
    \label{fig:meijerG}
\end{figure}

Using
\begin{align}
    & \quad \sum_{\sigma = \pm 1} \frac{1}{\sigma} \int_{-\sigma \infty}^{\theta_s} \mathrm{d}\theta_s'\,\langle \delta\hat\varphi \rangle_{\theta_f}(\theta_s') \exp\left(\frac{i \omega (\theta_s -\theta_s') }{\sigma \abs{v_\parallel} \langle \bhat \cdot\nabla\theta \rangle_{\theta_f}}  \right) \nonumber\\
    & = -\int_{ \infty}^{\theta_s} \mathrm{d}\theta_s'\,\langle \delta\hat\varphi \rangle_{\theta_f}(\theta_s') \exp\left(\frac{-i \omega (\theta_s -\theta_s') }{\abs{v_\parallel} \langle \bhat \cdot\nabla\theta \rangle_{\theta_f}}  \right) +\int_{ -\infty}^{\theta_s} \mathrm{d}\theta_s'\,\langle \delta\hat\varphi \rangle_{\theta_f}(\theta_s') \exp\left(\frac{i \omega (\theta_s -\theta_s') }{\abs{v_\parallel} \langle \bhat \cdot\nabla\theta \rangle_{\theta_f}}  \right) \nonumber \\
    & = \int_{-\infty}^{\infty}\mathrm{d}\theta_s' \,\langle \delta\hat\varphi \rangle_{\theta_f}(\theta_s') \exp\left(\frac{i \omega \abs{\theta_s -\theta_s'} }{\abs{v_\parallel} \langle \bhat \cdot\nabla\theta \rangle_{\theta_f}}  \right),
\end{align}
we derive the electron density fluctuation to be
\begin{equation} \label{eq:delta_ne_Pe}
    \frac{\delta \hat n_e}{n_e} = \frac{e \delta\hat\varphi}{T_e} + \frac{1}{n_e}\int\mathrm{d}^3v\,\hat h_{e,(0)} = \frac{e \delta\hat\varphi}{T_e}+\int_{-\infty}^\infty \mathrm{d}\theta_s' \frac{e \langle \delta\hat\varphi \rangle_{\theta_f}(\theta_s')}{T_e} \mathcal{P}_e(\theta_s-\theta_s') ,
\end{equation}
where we have introduced the electron propagator
\begin{equation} \label{eq:longtail_Pe}
    \mathcal{P}_e(\theta_s-\theta_s') = 2\pi i \int_0^\infty\mathrm{d}v_\parallel\int_0^\infty\mathrm{d}v_\perp v_\perp \, \frac{\omega - \tilde\omega_{*e}}{v_\parallel \langle \bhat\cdot\nabla\theta\rangle_{\theta_f}} \frac{F_{Me}}{n_e} \exp\left(\frac{i \omega \abs{\theta_s -\theta_s'} }{v_\parallel \langle \bhat \cdot\nabla\theta \rangle_{\theta_f}}  \right).
\end{equation}
Evaluating the integral over perpendicular velocities in \eqref{eq:longtail_Pe}, the electron propagator simplifies to
\begin{equation}
    \mathcal{P}_e(\theta_s-\theta_s') = \frac{i}{\sqrt{\pi}} \int_0^\infty \frac{\mathrm{d}v_\parallel}{v_{Te}}  \exp\left( - \frac{v_\parallel^2}{v_{Te}^2} + \frac{i \omega \abs{ \theta_s - \theta_s'}}{v_\parallel \langle \bhat\cdot\nabla\theta\rangle_{\theta_f}}  \right) \frac{\omega - \omega_{*e} - \omega_{*e}^T \left( \frac{v_\smallparallel^2}{v_{Te}^2} - \frac{1}{2} \right)}{v_\parallel \langle \bhat\cdot\nabla\theta \rangle_{\theta_f}}. \label{eq:longtail_Pe_vpa_integral}
\end{equation}
The parallel velocity integral may in turn be expressed in terms of Meijer G-functions \citep{meijer_uber_1936},
\begin{equation} \label{eq:Pe_MeijerG}
    \mathcal{P}_e(\theta_s-\theta_s') = \frac{i}{2\pi} \left( \frac{\omega - \omega_{*e}+\omega_{*e}^T/2}{\omega_{\parallel e}} \MeijerG[\bigg]{0}{0}{0}{3}{\dots\\}{0,0,1/2}{\frac{z_c^2}{4}} - \frac{\omega_{*e}^T}{\omega_{\parallel e}} \MeijerG[\bigg]{0}{0}{0}{3}{\dots\\}{0,1/2,1}{\frac{z_c^2}{4}} \right),
\end{equation}
where we defined the average transit frequency of an electron with thermal speed
\begin{equation} \label{eq:def_omegapare}
    \omega_{\parallel e} = v_{Te} \langle \bhat\cdot\nabla \theta \rangle_{\theta_f},
\end{equation}
and a rescaled parallel distance
\begin{equation} \label{eq:longtail_z_c}
    z_c = -i \frac{\omega}{\omega_{\parallel e}}  \abs{\theta_s-\theta_s'}.
\end{equation}
The Meijer G-functions required for the evaluation of the electron propagator \eqref{eq:Pe_MeijerG} are shown in Figure~\ref{fig:meijerG}. They are localised to $\abs{z_c} \sim 1$, i.e. the electron propagator's characteristic length scale in $\theta_s$ corresponds to the distance that a thermal electron travels within the mode timescale for growth or oscillation.

The electron propagator \eqref{eq:Pe_MeijerG} diverges logarithmically as $z_c \rightarrow 0$ because we did not take into account that particles with $v_\parallel \rightarrow 0$ become trapped or that they break the assumption of fast parallel streaming in the ordering \eqref{eq:longtail_multiscale_ordering}. However, the contribution from such particles to the fluctuating density is negligible in cases of interest: indeed, the integral \eqref{eq:delta_ne_Pe} over the electron propagator remains finite for any smoothly varying $\langle \delta\hat\varphi \rangle_{\theta_f}$.

\subsection{Dispersion relation}\label{sec:longtail_theory_dispersion}

Using expressions \eqref{eq:longtail_ni} for the ion density fluctuation and  \eqref{eq:delta_ne_Pe} for the electron density fluctuation, we can evaluate the quasineutrality equation \eqref{eq:QN_ballooning} to derive the dispersion relation for extended modes. Considering a single ion species for simplicity, we obtain
\begin{equation} \label{eq:quasineutrality_propagator}
    0  = -(1+\tau-\mathcal{M}_i) \delta\hat\varphi - \tau \int_{-\infty}^\infty \mathrm{d}\theta_s' \langle \delta\hat\varphi \rangle_{\theta_f}(\theta_s') \mathcal{P}_e(\theta_s-\theta_s'),
\end{equation}
where we used that $Z_i n_i = n_e$ by quasineutrality of the background, and we defined ${\tau = T_i/Z_i T_e}$. We rearrange equation \eqref{eq:quasineutrality_propagator} to
\begin{equation} \label{eq:tail_modes_disp_integral_eq_pre_integral}
    0 = \delta\hat\varphi + \frac{\tau}{1+\tau -\mathcal{M}_i} \int_{-\infty}^\infty \mathrm{d}\theta_s' \langle \delta\hat\varphi \rangle_{\theta_f}(\theta_s') \mathcal{P}_e(\theta_s-\theta_s') 
\end{equation}
and average over the fast coordinate $\theta_f$ to derive
\begin{equation} \label{eq:tail_modes_disp_integral_eq}
    0 = \mathcal{D}^\mathrm{extended}  = \langle \delta\hat\varphi \rangle_{\theta_f} + \left\langle \frac{\tau}{1+\tau -\mathcal{M}_i} \right\rangle_{\theta_f} \int_{-\infty}^\infty \mathrm{d}\theta_s' \langle \delta\hat\varphi \rangle_{\theta_f}(\theta_s') \mathcal{P}_e(\theta_s-\theta_s').
\end{equation}
This integral equation for $\langle \delta\hat\varphi \rangle_{\theta_f}(\theta_s)$ is the dispersion relation for extended tail modes at low magnetic shear. For a given solution to \eqref{eq:tail_modes_disp_integral_eq}, the potential fluctuation's variation in $\theta_f$ may be recovered using 
\begin{equation} \label{eq:local_delta_phihat}
    \delta\hat\varphi = \frac{\tau}{1+\tau -\mathcal{M}_i} \left\langle \frac{\tau}{1+\tau -\mathcal{M}_i} \right\rangle_{\theta_f}^{-1} \langle \delta\hat\varphi \rangle_{\theta_f},
\end{equation}
which follows from replacing \eqref{eq:tail_modes_disp_integral_eq} in \eqref{eq:tail_modes_disp_integral_eq_pre_integral}.

We note that, when the electron propagation over the scale of the mode is very fast compared to the mode frequency and growth rate, i.e. $\omega_{\parallel e} / \abs{\theta_s} \gg \abs{\omega}$, the integral of such a mode multiplied by the electron propagator $\mathcal{P}_e$ may be neglected in \eqref{eq:quasineutrality_propagator} and we obtain instead a dispersion relation for localised ion-scale modes
\begin{equation} \label{eq:D_localised}
    0 = \mathcal{D}^\mathrm{localised} = 1+\tau - \mathcal{M}_i.
\end{equation}
While the electron propagator still determines the localised mode's eigenfunction in $\theta$, the localised mode's frequency is independent of the propagator to leading order. The dispersion relation \eqref{eq:D_localised} describes toroidal ITG modes \citep{horton_toroidal_1981, guzdar_ion-temperature-gradient_1983, biglari_toroidal_1989, zocco_threshold_2018} in the limit of negligible ion parallel streaming.

\subsection{Slow electron propagation limit}\label{sec:longtail_theory_weak_limit}

An instructive limit of the extended tail mode theory may be obtained by considering the mode extent along the magnetic field to be large compared with the typical distance of electron propagation within an oscillation period. As we will discuss in Section~\ref{sec:longtail_conditions_mode_width}, this limit of slow electron propagation is typically satisfied for the fastest growing GEMs because the parallel streaming of electrons stabilises the GEM. We note that the electrons are still assumed to propagate quickly over the distance $\theta_f \sim 1$.

In the slow electron propagation limit, the electron propagator $\mathcal{P}_e$ decays in $\theta_s$ over a distance that is short compared to the mode width. We may therefore Taylor expand the potential fluctuation in the electron density response \eqref{eq:delta_ne_Pe}: to the order required, we find
\begin{align} \label{eq:delta_ne_Pe_weak}
     & \int_{-\infty}^\infty  \mathrm{d}\theta_s' \frac{e \langle \delta\hat\varphi \rangle_{\theta_f}(\theta_s')}{T_e} \mathcal{P}_e(\theta_s-\theta_s')  \\
      \approx \frac{e}{T_e} & \int_{-\infty}^\infty \mathrm{d}\theta_s' \left( \langle \delta\hat\varphi \rangle_{\theta_f}\bigg|_{\theta_s} + \dd{\langle \delta\hat\varphi \rangle_{\theta_f}}{\theta_s}\bigg|_{\theta_s} (\theta_s'-\theta_s) + \dd{^2 \langle \delta\hat\varphi \rangle_{\theta_f}}{\theta_s^2}\bigg|_{\theta_s} \frac{(\theta_s'-\theta_s)^2}{2}  \right) \mathcal{P}_e(\theta_s-\theta_s').\nonumber
\end{align}
The required integrals over the electron propagator may be derived directly from \eqref{eq:longtail_Pe_vpa_integral},
\begin{align}
    &\int_{-\infty}^\infty\mathrm{d}x \,\mathcal{P}_e(x) = -1 + \frac{\omega_{*e}}{\omega}, \label{eq:longtail_integral_over_Pe}\\
    &\int_{-\infty}^\infty\mathrm{d}x \,\mathcal{P}_e(x) x = 0,\\
    &\int_{-\infty}^\infty\mathrm{d}x \,\mathcal{P}_e(x) \frac{x^2}{2} = \frac{\omega_{\parallel e}^2}{2\omega^2} \left( 1 - \frac{\omega_{*e}}{\omega} - \frac{\omega_{*e}^T}{\omega}\right).
\end{align}
The electron density fluctuation \eqref{eq:delta_ne_Pe} then simplifies to
\begin{equation}\label{eq:delta_ne_total_Pe_weak}
    \frac{\delta \hat n_e}{n_e} = \frac{ e}{T_e} \left( \delta\hat\varphi - \langle \delta\hat\varphi \rangle_{\theta_f}  \right) + \frac{\omega_{*e}}{\omega} \frac{ e \langle \delta\hat\varphi \rangle_{\theta_f}}{T_e} + \frac{e}{T_e} \dd{^2 \langle \delta\hat\varphi \rangle_{\theta_f}}{\theta_s^2} \frac{\omega_{\parallel e}^2}{2\omega^2} \left( 1- \frac{\omega_{*e}+\omega_{*e}^T}{\omega} \right).
\end{equation}
In the limit of slow electron propagation, the integral dispersion relation \eqref{eq:tail_modes_disp_integral_eq} becomes a second order differential equation
\begin{align}
    \langle \delta\hat\varphi \rangle_{\theta_f} = & \left\langle \frac{\tau}{1+\tau -\mathcal{M}_i} \right\rangle_{\theta_f} \bigg[  \left(1 - \frac{\omega_{*e}}{\omega} \right) \langle \delta\hat\varphi \rangle_{\theta_f} \nonumber\\
     & - \frac{\omega_{\parallel e}^2}{2\omega^2}\left( 1 - \frac{\omega_{*e}}{\omega} - \frac{\omega_{*e}^T}{\omega}\right) \dd{^2 \langle \delta\hat\varphi \rangle_{\theta_f}}{\theta_s^2} \bigg],
\end{align}
or equivalently,
\begin{align} \label{eq:longtail_disp_rel_weak_Pe}
    0 = \mathcal{D}^{ \mathrm{slow}} = & \frac{\omega_{\parallel e}^2}{2\omega^2} \left( 1 - \frac{\omega_{*e}}{\omega} - \frac{\omega_{*e}^T}{\omega}\right) \dd{^2 \langle \delta\hat\varphi \rangle_{\theta_f}}{\theta_s^2} \nonumber\\
     & +  \left( \left\langle \frac{\tau}{1+\tau -\mathcal{M}_i} \right\rangle_{\theta_f}^{-1} - \left( 1 - \frac{\omega_{*e}}{\omega} \right) \right) \langle \delta\hat\varphi \rangle_{\theta_f}.
\end{align}
Equation \eqref{eq:longtail_disp_rel_weak_Pe} is the dispersion relation of extended tail modes in the slow electron propagation limit. We will find numerically in Section~\ref{sec:longtail_sims} that this limit is appropriate to describe the geodesic extended modes.

\subsection{Fluid ion limit} \label{sec:longtail_model_disp}

It will prove instructive to further consider the dispersion relation \eqref{eq:longtail_disp_rel_weak_Pe} in the `fluid' limit of non-resonant ions and long perpendicular wavelengths, corresponding to the subsidiary ordering
\begin{equation}
    k_\perp \rho_i \sim \frac{\tilde{\omega}_{Mi}}{\omega} \ll \frac{\tilde \omega_{*i}}{\omega} \sim 1.
\end{equation}
We may then approximate the ion response function \eqref{eq:longtail_Mi} as
\begin{align} \label{eq:M_i_fluid}
    \mathcal{M}_i \approx &
    \; 1-\frac{\omega_{*i}}{\omega} - \frac{k_\perp^2 \rho_i^2}{2} \left( 1 - \frac{\omega_{*i}+\omega_{*i}^T}{\omega} \right) \nonumber\\
    & + \frac{\omega_{Mi}}{\omega} \left( 1 - \frac{\omega_{*i}+\omega_{*i}^T}{\omega} \right) + \frac{7}{4}\frac{\omega_{Mi}^2}{\omega^2}  \left( 1 - \frac{\omega_{*i}+2\omega_{*i}^T}{\omega} \right),
\end{align}
where we assumed $\omega_{\nabla B,i} = \omega_{\kappa,i}=\omega_{Mi}$, which is satisfied for small plasma $\beta \ll 1$ or for small binormal wavenumbers $k_y \ll k_x$ (this is automatically satisfied for $\hat s\theta_s \gg 1$, see \eqref{eq:kperp_tail}). 

The corresponding contribution from \eqref{eq:M_i_fluid} to the dispersion relation \eqref{eq:longtail_disp_rel_weak_Pe} is
\begin{align}
    & \left\langle \frac{\tau}{1+\tau -\mathcal{M}_i} \right\rangle_{\theta_f}^{-1}  \approx 1 + \frac{\omega_{*i}}{\tau\omega} +  \frac{\langle k_\perp^2 \rho_i^2 \rangle_{\theta_f}}{2 \tau} \left( 1 - \frac{\omega_{*i}+\omega_{*i}^T}{\omega} \right) \nonumber\\
    & \qquad - \frac{\langle \omega_{Mi}^2 \rangle_{\theta_f}}{\tau\omega^2} \left[ \frac{7}{4} \left( 1 - \frac{\omega_{*i}+2\omega_{*i}^T}{\omega} \right) + \left( \tau + \frac{\omega_{*i}}{\omega} \right)^{-1} \left( 1 - \frac{\omega_{*i}+\omega_{*i}^T}{\omega} \right)^2 \right] ,
\end{align}
where we assumed $\langle \omega_{Mi} \rangle_{\theta_f} = 0$; we will discuss this assumption in Section~\ref{sec:longtail_limits_validity}. For sufficiently large $\theta_s$, the averaged finite Larmor radius effect and magnetic drift terms may be approximated as
\begin{alignat}{2}
    \frac{\langle k_\perp^2 \rho_i^2 \rangle_{\theta_f}}{2} & \approx \overline{b}_{i0} (C_b + \hat s^2 \theta_s^2), \\
    \langle \omega_{Mi}^2\rangle_{\theta_f} & \approx  \frac{\overline{\omega}_{Mi0}^2}{2} (C_M + \hat s^2 \theta_s^2),
\end{alignat}
with constant coefficients 
\begin{align}
    \overline{b}_{i0} & = \frac{k_y^2 \rho_i^2}{2} \left( \DD{y}{\alpha} \right)^2 \left( \frac{\psi_0}{q(\psi_0)} \DD{x}{\psi} \right)^{-2} \left\langle \abs{\nabla x}^2 \right\rangle_{\theta_f}, \\
    \overline{\omega}_{Mi0}^2 & = 2 k_y^2 v_{Ti}^4 \left( \DD{y}{\alpha} \right)^2 \left( \frac{\psi_0}{q(\psi_0)} \DD{x}{\psi} \right)^{-2} \left\langle \left(\frac{\nabla x \cdot \bhat \times \nabla \ln B}{\Omega_i} \right)^2 \right\rangle_{\theta_f},
\end{align}
and $C_b>0$, $C_M>0$. In a large aspect ratio circular cross-section tokamak, $C_b = C_M =1$, which we will now consider for simplicity. We may then write the dispersion relation \eqref{eq:longtail_disp_rel_weak_Pe} as 
\begin{align} \label{eq:longtail_disp_rel_weak_Pe_NR_ions}
    0 & = \frac{\omega_{\parallel e}^2}{2\omega^2} \left( 1 - \frac{\omega_{*e}}{\omega} - \frac{\omega_{*e}^T}{\omega}\right) \dd{^2 \langle \delta\hat\varphi \rangle_{\theta_f}}{\theta_s^2}  + \Bigg\{ \frac{\overline{b}_{i0}}{\tau} \left( 1 - \frac{\omega_{*i}+\omega_{*i}^T}{\omega} \right) \\\nonumber
    & - \frac{\overline{\omega}_{Mi0}^2}{2 \tau \omega^2} \left[ \frac{7}{4} \left( 1 - \frac{\omega_{*i}+2\omega_{*i}^T}{\omega} \right) + \left( \tau + \frac{\omega_{*i}}{\omega} \right)^{-1} \left( 1 - \frac{\omega_{*i}+\omega_{*i}^T}{\omega} \right)^2 \right] \Bigg\} (1+\hat s^2 \theta_s^2) \langle \delta\hat\varphi \rangle_{\theta_f},
\end{align}
where we used $\omega_{*i}+\tau\omega_{*e}=0$, which follows from the quasineutrality of the background.

Equation \eqref{eq:longtail_disp_rel_weak_Pe_NR_ions} is equivalent to a quantum harmonic oscillator with negative energy and for this reason, bounded eigenfunctions do not exist to lowest order. Indeed, the bound solutions of a quantum harmonic oscillator are well known, as are their energy levels: these are quantised and strictly positive. We illustrate the problem by inserting the quantum harmonic oscillator ground state solution
\begin{equation}
    \langle \delta\hat\varphi \rangle_{\theta_f} = \delta\hat\varphi_0 \exp(-A\theta_s^2)
\end{equation}
into \eqref{eq:longtail_disp_rel_weak_Pe_NR_ions}. We find either an exponentially growing solution in $\abs{\theta}$ (i.e. Re$(A) < 0$), or
\begin{align}\label{eq:longtail_disp_rel_weak_Pe_NR_ions_approximate}
    0 &= \mathcal{D}^\mathrm{fluid} = \frac{\overline{b}_{i0}}{\tau} \left( 1 - \frac{\omega_{*i}+\omega_{*i}^T}{\omega} \right) \nonumber\\
    & - \frac{\overline{\omega}_{Mi0}^2}{2 \tau \omega^2} \left[ \frac{7}{4} \left( 1 - \frac{\omega_{*i}+2\omega_{*i}^T}{\omega} \right) + \left( \tau + \frac{\omega_{*i}}{\omega} \right)^{-1} \left( 1 - \frac{\omega_{*i}+\omega_{*i}^T}{\omega} \right)^2 \right].
\end{align}
In the latter case, $A = 0$, such that the eigenmode is constant in $\theta_s$ instead of decaying at large $\abs{\theta_s}$. The non-resonant expansion carried out in this section cannot therefore capture the correct eigenfunction of the extended modes: higher order contributions in $k_\perp \rho_i$ and $\omega_{Mi}/\omega$ must be included in the expansion of the ion response function \eqref{eq:longtail_Mi}. These contributions could originate from higher order non-resonant corrections or from resonant corrections; we leave the study of such terms to future work.

Even though these small corrections to the ion response function are required to determine the eigenmode, they are not necessarily important for the determination of the mode frequency. Indeed, \eqref{eq:longtail_disp_rel_weak_Pe_NR_ions} is well-behaved when the electron parallel streaming becomes negligible ($\omega_{\parallel e}\rightarrow 0$), in which case the fluid dispersion relation \eqref{eq:longtail_disp_rel_weak_Pe_NR_ions_approximate} approximately holds. We will show in Section~\ref{sec:longtail_sims} that the fluid dispersion relation recovers the GEM frequency as the electron propagation over the mode extent becomes negligibly slow. Due to its simplicity, the dispersion relation \eqref{eq:longtail_disp_rel_weak_Pe_NR_ions_approximate} will prove useful to understand the physics of the GEMs, at least qualitatively (see discussion in Section~\ref{sec:longtail_physics_nonresonant}).

\section{Simulations of extended modes}\label{sec:longtail_sims}

We first discuss in Section~\ref{sec:longtail_sims_numerical} our algorithm to solve numerically the derived dispersion relations, as well as various numerical parameters. We then verify our theory against \texttt{stella} \citep{barnes_stella_2019} gyrokinetic simulations in Section~\ref{sec:longtail_param_scans} and consider various parameter dependences of the extended modes.

\subsection{Numerical methods and parameters}\label{sec:longtail_sims_numerical}

To find the mode frequency $\omega$ and the fluctuating potential $\langle \delta\hat\varphi \rangle_{\theta_f}$ which solve the extended tail mode dispersion relation \eqref{eq:tail_modes_disp_integral_eq}, we consider the eigenvalue problem
\begin{equation} \label{eq:tail_modes_eigenvalue_problem}
     \lambda \langle \delta\hat\varphi \rangle_{\theta_f} = - \left\langle \frac{\tau}{1+\tau -\mathcal{M}_i} \right\rangle_{\theta_f} \int_{-\infty}^\infty \mathrm{d}\theta_s' \langle \delta\hat\varphi \rangle_{\theta_f}(\theta_s') \mathcal{P}_e(\theta_s-\theta_s'),
\end{equation}
which may be solved for a given value of $\omega$. By using a root finding algorithm to minimise $\abs{\lambda -1}$, we obtain the $\omega$ and $\langle \delta\hat\varphi \rangle_{\theta_f}$ which solve the dispersion relation \eqref{eq:tail_modes_disp_integral_eq}. 

The eigenvalue problem \eqref{eq:tail_modes_eigenvalue_problem} is solved on a uniform grid
\begin{equation}
    \theta_{s,n} = \frac{\theta_M}{2} \left(-1  + \frac{2n}{N_\theta-1}\right),
\end{equation}
with $n \in [0,1,...,N_\theta-1]$. The tube length $\theta_M$ and resolution $N_\theta$ are inputs. The eigenvalue equation \eqref{eq:tail_modes_eigenvalue_problem} is then discretised as
\begin{equation}  \label{eq:tail_modes_eigenvalue_problem_discr}
    \lambda \Phi_i = \sum_j A_{ij} \Phi_j,
\end{equation}
where
\begin{equation}
    \Phi_i = \langle \delta\hat\varphi \rangle_{\theta_f}(\theta_{s,i})
\end{equation}
and
\begin{equation} \label{eq:longtail_Aij}
    A_{ij} = - \left\langle \frac{\tau}{1+\tau -\mathcal{M}_i} \right\rangle_{\theta_f} (\theta_{s,i})  \; \mathcal{P}_e(\theta_{s,i}-\theta_{s,j}) \Delta\theta_s,
\end{equation}
with the grid spacing $\Delta\theta_s = \theta_M/(N_\theta-1)$. An inverse power method \citep[see e.g.][]{franklin_computational_2013} is used to find the eigenvalue of \eqref{eq:tail_modes_eigenvalue_problem_discr} closest to $\lambda = 1$. 

When the electron propagator $\mathcal{P}_e$ in \eqref{eq:longtail_Aij} is evaluated through Meijer G-functions \eqref{eq:Pe_MeijerG}, we introduce an additional normalisation factor in $A_{ij}$ to ensure that the identity \eqref{eq:longtail_integral_over_Pe} is exactly satisfied numerically. Instead of \eqref{eq:longtail_Aij}, we thus use
\begin{equation}
    A_{ij} = - \left\langle \frac{\tau}{1+\tau -\mathcal{M}_i} \right\rangle_{\theta_f} (\theta_{s,i})  \; \mathcal{P}_e(\theta_{s,i}-\theta_{s,j}) \Delta\theta_s \frac{\omega_{*e}/\omega-1}{\int_{-\infty}^\infty \mathrm{d}\theta_s \mathcal{P}_e(\theta_s)},
\end{equation}
with the infinite integral in $\theta_s$ evaluated as
\begin{equation}
    \int_{-\infty}^\infty \mathrm{d}\theta_s \mathcal{P}_e(\theta_s) \rightarrow \Delta\theta_s \sum_i \mathcal{P}_e(\theta_{s,i}).
\end{equation}
This normalisation factor helps to avoid numerical instabilities which occur even for small numerical errors in the integral \eqref{eq:longtail_integral_over_Pe} and would therefore force us to consider computationally intractable values of $N_\theta$ and $\theta_M$.

In the slow electron propagation limit considered in Section~\ref{sec:longtail_theory_weak_limit}, the electron propagator in \eqref{eq:longtail_Aij} is instead evaluated as
\begin{equation}
    \mathcal{P}_e^\mathrm{slow}(\theta_{s,i}-\theta_{s,j}) = -\left(1 - \frac{\omega_{*e}}{\omega} \right)\delta_{i,j} + \frac{\omega_{\parallel e}^2}{2\omega^2}\left( 1 - \frac{\omega_{*e}}{\omega} - \frac{\omega_{*e}^T}{\omega}\right) D_{i,j} ,
\end{equation}
where $\delta_{i,j}$ is the Kronecker delta, and 
\begin{equation}
    D_{i,j} = \frac{1}{(\Delta\theta_s)^2}\cdot
    \begin{cases}
        \delta_{i,j+2} - 2\delta_{i,j+1} + \delta_{i,j} \qquad & (j=0),\\
        \delta_{i,j+1} - 2\delta_{i,j} + \delta_{i,j-1} \qquad & (0 < j < N_\theta-1), \\
        \delta_{i,j} - 2\delta_{i,j-1} + \delta_{i,j-2} \qquad & (j=N_\theta-1),
    \end{cases}
\end{equation}
is the discretised second-order derivative operator. Equation \eqref{eq:longtail_integral_over_Pe} is exactly satisfied numerically in this case. 

Most results we present below assume a tube length corresponding to $\theta^M / (2\pi) = 101$ poloidal turns and a parallel resolution of $N_\theta / 101 = 32$ grid points per poloidal turn. For the simulations at small electron mass $m_e/m_i \leq 1/3672$ or small magnetic shear $\hat s \leq 0.025$, we consider longer tubes  $\theta^M / (2\pi) = 201$, keeping the same number of grid points per poloidal turn. The same tube lengths and resolution per turn are used in the gyrokinetic simulations.

We use the \texttt{stella} code \citep{barnes_stella_2019} to perform gyrokinetic simulations in a flux tube of a tokamak with circular flux surfaces and aspect ratio $R/a=6$, and we consider the flux surface at $r/a = 0.3$. Here, $r$ is the radial distance from the magnetic axis to the circular flux surface under consideration, and $a$ is its value of $r$ for the outermost flux surface. We perform linear simulations to find the fastest growing mode for a given binormal wavenumber $k_y$ and with $\theta_0=0$ in all cases, without loss of generality. Indeed, the extended modes are very weakly dependent on $\theta_0$ as its contribution to the perpendicular wavenumber is negligible compared with $\theta_s$ for extended modes, see equation \eqref{eq:kperp_tail} and the discussion in \cite{hardman_extended_2022}. In all simulations, the velocity-space resolution is $N_{v_\smallparallel} = 64$, $N_\mu = 16$, max$(v_\parallel)=3.5\, v_{Ti}$, and the maximum of $\mu$ is chosen such that the maximum perpendicular velocity is $3 v_{Ti}$ at the minimum value of $B$. The ion charge is taken to be $Z_i = 1$ and the electron and ion temperatures are set to be equal. The plasma beta is set to $\beta=0$, such that $\omegakappai = \omeganablaBi$ in the simulations.

\subsection{Parameter scans} \label{sec:longtail_param_scans}

\begin{figure}
    \centering
    \includegraphics[width=\textwidth]{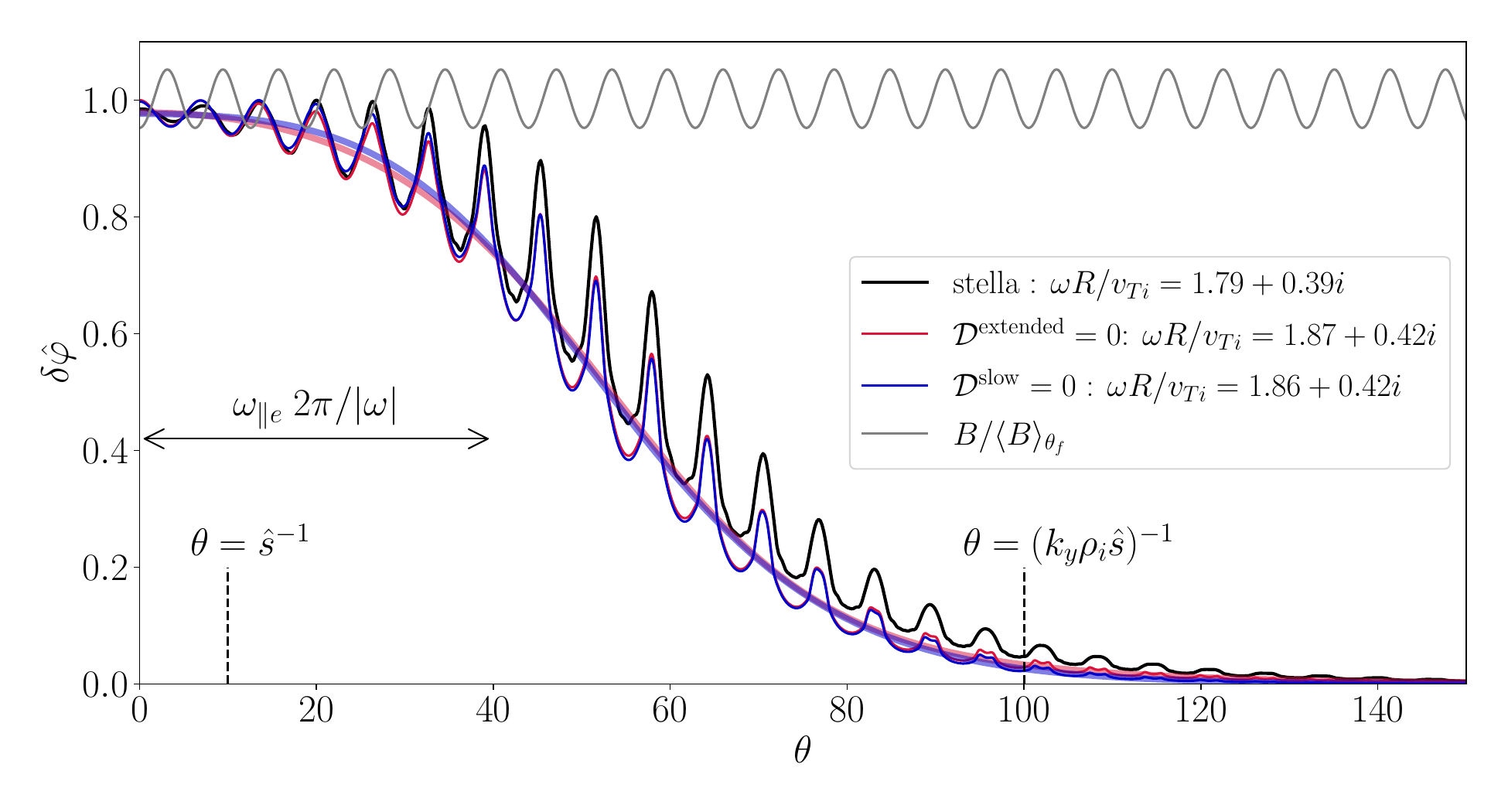}
    \caption[Extended tail eigenmode from gyrokinetic simulations and dispersion relation]{Extended tail eigenmode obtained from \texttt{stella} \citep{barnes_stella_2019} gyrokinetic simulations (black curve) and from the dispersion relation \eqref{eq:tail_modes_disp_integral_eq}, for the general electron propagator \eqref{eq:Pe_MeijerG} (red curves) and for the slow electron propagation limit (blue curves). The thick transparent curves are the averages $\langle \delta\hat\varphi\rangle_{\theta_f}$ of the eigenmode in the fast coordinate, found as solutions to the dispersion relation \eqref{eq:tail_modes_disp_integral_eq}. The variation of $B$ along the magnetic field is plotted for reference (grey curve) and the mode frequencies are provided in the legend. We only show the eigenmode for positive values of $\theta$ as it is symmetric in $\theta$. We model a Deuterium plasma ($\sqrt{m_i/m_e} = 60.6, Z_i = 1$) on the flux surface $r/a=0.3$ of a tokamak with circular cross-section, aspect ratio $R/a = 6$, safety factor $q=5$, and magnetic shear $\hat s = 0.1$. We consider a binormal wavenumber $k_y \rho_i = 0.1$ and a pure ion-temperature-gradient drive $a/L_{Ti}=4$ and $a/L_n = a/L_{Te}=0$. The vertical black dashed lines indicate the distance along the field line where the radial and binormal wavenumbers are of similar size ($\theta = \hat s^{-1} \Leftrightarrow  k_x \approx k_y $) and where the ion FLR effects become significant ($\theta = (k_y \rho_i \hat s)^{-1} \Leftrightarrow  k_\perp \rho_i \approx 1$). The horizontal left-right arrow gives the distance of parallel propagation of a thermal electron within the characteristic timescale of the mode ($\Delta\theta = \omega_{\parallel e} 2\pi/\abs{\omega}$).}
    \label{fig:longtail_phi_theta_ref_case}
\end{figure}

In Figure~\ref{fig:longtail_phi_theta_ref_case}, we show that the dispersion relation \eqref{eq:tail_modes_disp_integral_eq} captures the shape and frequency of the extended mode obtained from gyrokinetic simulations with good accuracy. Furthermore, the slow electron propagation limit presented in Section~\ref{sec:longtail_theory_weak_limit} proves adequate in this case, as it provides an eigenmode that is nearly identical to the solution of the dispersion relation using the more general propagator \eqref{eq:Pe_MeijerG}.

We take the simulation of Figure~\ref{fig:longtail_phi_theta_ref_case} to be our reference case and study how varying several parameters affects the mode frequency and eigenfunction, as well as the accuracy of the dispersion relations \eqref{eq:tail_modes_disp_integral_eq} and \eqref{eq:longtail_disp_rel_weak_Pe}. We also include in our comparisons the complex frequency of the fastest growing mode from the fluid dispersion relation \eqref{eq:longtail_disp_rel_weak_Pe_NR_ions_approximate}.

\subsubsection{Varying magnetic shear}

\begin{figure}
    \centering
            \includegraphics[width=0.8\textwidth, trim={0.6cm 0.8cm 0.4cm 0.0cm}, clip]{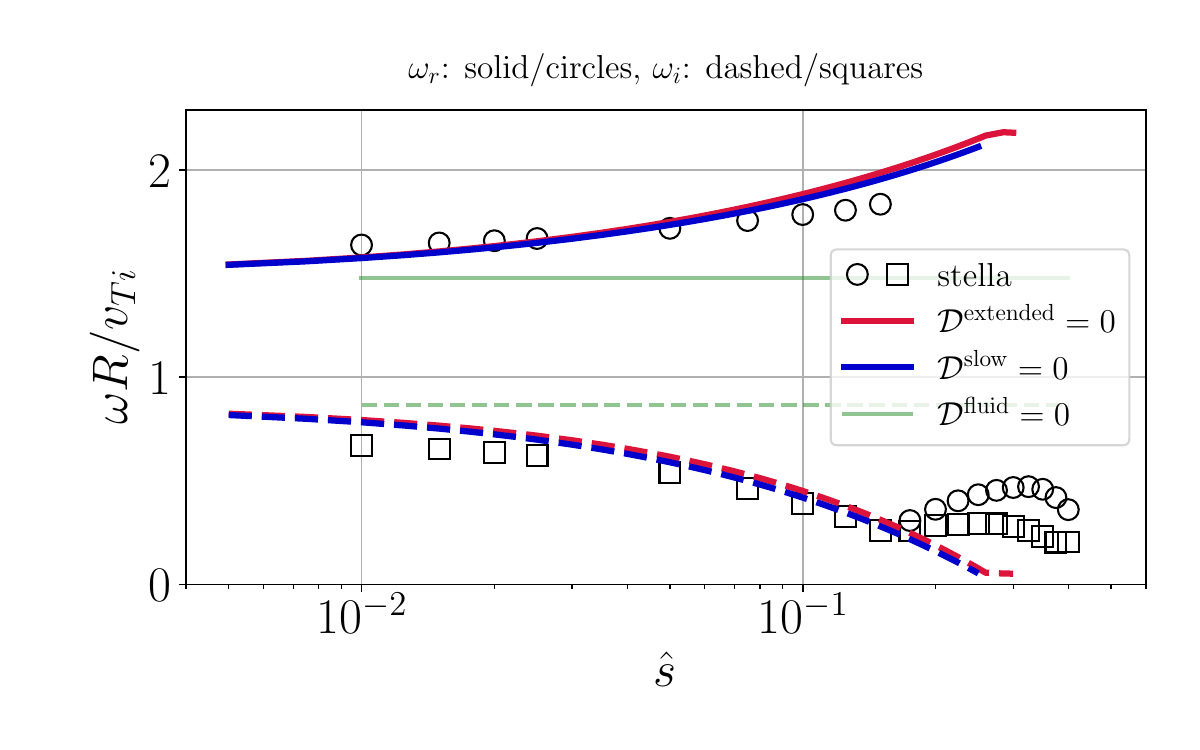}
    \caption[Extended mode frequency for varying magnetic shear]{Complex frequency for varying magnetic shear -- other parameters are identical to the reference case of Figure~\ref{fig:longtail_phi_theta_ref_case}. The solid and dashed curves correspond to the real and imaginary frequencies, respectively, obtained from solving the dispersion relations \eqref{eq:tail_modes_disp_integral_eq} (red), \eqref{eq:longtail_disp_rel_weak_Pe} (blue), and \eqref{eq:longtail_disp_rel_weak_Pe_NR_ions_approximate} (green). The circles and squares correspond to the real and imaginary frequencies, respectively, obtained from gyrokinetic simulations.}
    \label{fig:tailmode_omega_shat}
\end{figure}

First, we consider the effect of varying the magnetic shear in Figure~\ref{fig:tailmode_omega_shat}. Increasing the magnetic shear reduces the growth rate, i.e. it stabilises the extended modes. Localised modes with smaller real frequency become dominant for $\hat s \gtrsim 0.2$; these are captured by the gyrokinetic simulations but not by the theory for extended modes. For $\hat s \lesssim 0.2$, the dispersion relations \eqref{eq:tail_modes_disp_integral_eq} and \eqref{eq:longtail_disp_rel_weak_Pe} capture the complex mode frequency accurately. For very small values of the magnetic shear, the eigenmodes become very extended and the solution tends towards the fluid dispersion relation \eqref{eq:longtail_disp_rel_weak_Pe_NR_ions_approximate}.

We note that the extended modes have a large real frequency $\omega_r \sim v_{Ti}/R > 0$, corresponding to propagation in the ion diamagnetic direction. These are the rapidly oscillating geodesic extended modes (GEMs), whose fast oscillation will be explained in Section~\ref{sec:longtail_discussion} below. In contrast, the localised modes have a small real frequency, as shown by the gyrokinetic simulations for $\hat s \gtrsim 0.2$ in Figure~\ref{fig:tailmode_omega_shat}.

\subsubsection{Varying electron mass and safety factor}

\begin{figure}
    \centering
    \begin{subfigure}[t]{0.49\columnwidth}
            \centering
            \includegraphics[width=\textwidth, trim={0.6cm 0.8cm 0.4cm 0cm}, clip]{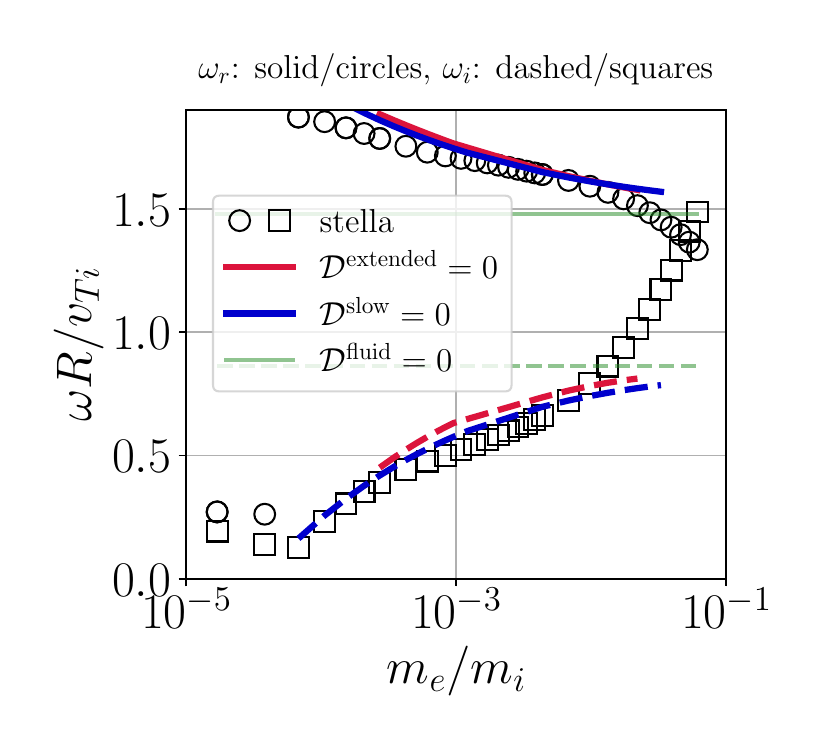}
        \caption{Electron mass scan}
        \label{fig:tailmode_omega_me}
    \end{subfigure}
   \begin{subfigure}[t]{0.49\columnwidth}
            \centering
            \includegraphics[width=\textwidth, trim={0.6cm 0.8cm 0.4cm 0cm}, clip]{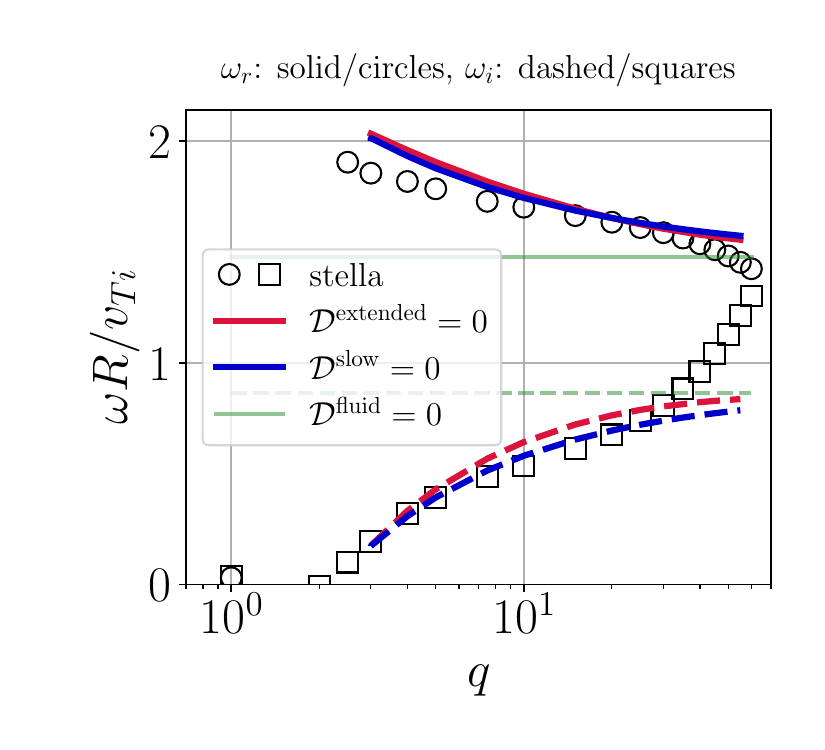}
        \caption{Safety factor scan}
        \label{fig:tailmode_omega_qinp}
    \end{subfigure}
    
    \caption[Extended mode frequency for varying electron mass and safety factor]{Complex frequency for varying electron mass (a) and safety factor (b) -- other parameters are identical to the reference case of Figure~\ref{fig:longtail_phi_theta_ref_case}. The solid and dashed curves correspond to the real and imaginary frequencies, respectively, obtained from solving the dispersion relations \eqref{eq:tail_modes_disp_integral_eq} (red), \eqref{eq:longtail_disp_rel_weak_Pe} (blue), and \eqref{eq:longtail_disp_rel_weak_Pe_NR_ions_approximate} (green). The circles and squares correspond to the real and imaginary frequencies, respectively, obtained from gyrokinetic simulations.}
    \label{fig:tailmode_omega_qinp_me}
\end{figure}

Next, we consider the effects of varying the electron mass and safety factor in Figure~\ref{fig:tailmode_omega_qinp_me}. We note that $\bhat \cdot \nabla \theta = (qR)^{-1}$ in the circular cross-section tokamak considered here, such that the characteristic electron parallel propagation frequency \eqref{eq:def_omegapare} is given by
\begin{equation}
    \omega_{\parallel e} = \frac{v_{Te}}{qR} \sim  \frac{v_{Ti}}{R} \frac{1}{q} \sqrt{\frac{m_i}{m_e}},
\end{equation}
assuming $T_e \sim T_i$. Because we ignored the ion parallel streaming in Section~\ref{sec:longtail_theory_ions}, the electron-to-ion mass ratio and the safety factor only enter the extended mode theory through their combination in $\omega_{\parallel e}$, explaining the similarities between Figures~\ref{fig:tailmode_omega_me} and \ref{fig:tailmode_omega_qinp}.

As $m_e/m_i$ or $q$ are reduced, $\omega_{\| e}/\omega$ increases, making the non-adiabatic electron response more important (see the last term in \eqref{eq:delta_ne_total_Pe_weak}) and thus stabilising the extended modes. Eventually, for very small $m_e/m_i \lesssim 10^{-4}$ or $q \lesssim 2$, the extended modes become subdominant to localised ion-scale modes with smaller real frequency. Oppositely, as $m_e/m_i$ or $q$ are increased, the stabilising electron parallel streaming becomes progressively weaker, the GEM growth rate increases, and the fluid dispersion relation \eqref{eq:longtail_disp_rel_weak_Pe_NR_ions_approximate} becomes more accurate. However, for very large $m_e/m_i \gtrsim 10^{-2}$ or $q \gtrsim 40$, the dispersion relations \eqref{eq:tail_modes_disp_integral_eq} and \eqref{eq:longtail_disp_rel_weak_Pe} fail to capture the frequency from gyrokinetic simulations because the electron propagation becomes so slow that the electrons can no longer be assumed to propagate quickly over $\theta_f\sim 1$ scales.

\subsubsection{Varying binormal wavenumber and ion temperature gradient}

\begin{figure}
    \centering

   \begin{subfigure}[t]{0.49\columnwidth}
            \centering
            \includegraphics[width=\textwidth, trim={0.6cm 0.8cm 0.4cm 0.0cm}, clip]{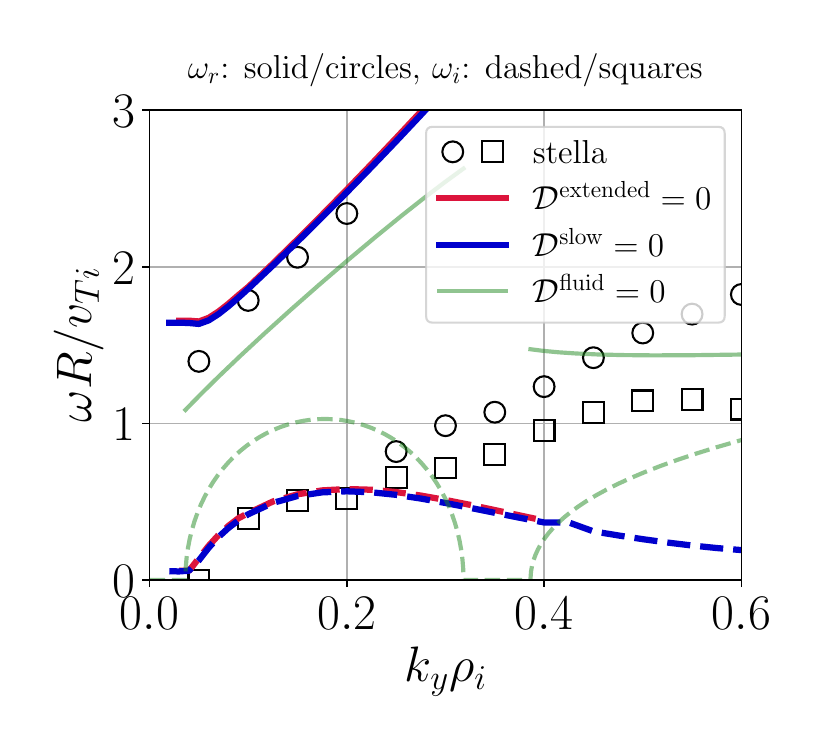}
        \caption{Binormal wavenumber scan}
        \label{fig:tailmode_omega_ky}
    \end{subfigure}
    \begin{subfigure}[t]{0.49\columnwidth}
            \centering
            \includegraphics[width=\textwidth, trim={0.6cm 0.8cm 0.4cm 0.0cm}, clip]{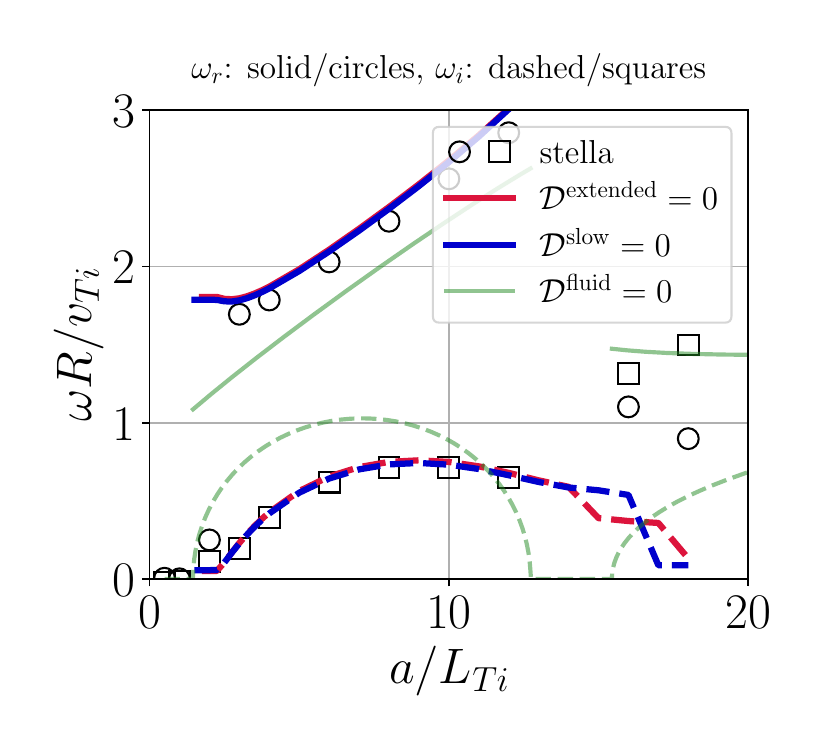}
        \caption{Ion temperature gradient scan}
        \label{fig:tailmode_omega_tiprim}
    \end{subfigure}

    \caption[Extended mode frequency for varying binormal wavenumber and ion temperature gradient]{Complex mode frequency for varying binormal wavenumber (a) and ion temperature gradient (b) -- other parameters are identical to the reference case of Figure~\ref{fig:longtail_phi_theta_ref_case}. The solid and dashed curves correspond to the real and imaginary frequencies, respectively, obtained from solving the dispersion relations \eqref{eq:tail_modes_disp_integral_eq} (red), \eqref{eq:longtail_disp_rel_weak_Pe} (blue), and \eqref{eq:longtail_disp_rel_weak_Pe_NR_ions_approximate} (green). The circles and squares correspond to the real and imaginary frequencies, respectively, obtained from gyrokinetic simulations.}
    \label{fig:tailmode_omega_ky_tiprim}
\end{figure}

In Figure~\ref{fig:tailmode_omega_ky_tiprim}, we show the mode frequency as a function of the binormal wavenumber and the ion temperature gradient drive.  The extended modes are the fastest growing modes at small $k_y \rho_i \lesssim 0.2$ and $a/L_{Ti} \lesssim 15$. For larger $k_y \rho_i \gtrsim 0.2$ or large ion temperature gradient values $a/L_{Ti} \gtrsim 15$, localised modes with smaller real frequency become dominant in the gyrokinetic simulations. The fluid dispersion relation \eqref{eq:longtail_disp_rel_weak_Pe_NR_ions_approximate} captures these trends qualitatively, though it fails to quantitatively predict the correct mode frequency because the stabilising electron parallel streaming is not negligible for the chosen values of $q$, $m_e/m_i$, and $\hat s$.

Figure~\ref{fig:tailmode_omega_ky} shows that the GEMs have a finite real frequency as the binormal wavenumber is made vanishingly small ($k_y \rightarrow 0$), though their growth rate vanishes below some wavenumber threshold. The GEMs also have an instability threshold in the ion temperature gradient, as evidenced by Figure~\ref{fig:tailmode_omega_tiprim}.

\subsubsection{Varying electron temperature gradient and density gradient}

\begin{figure}
    \centering
   \begin{subfigure}[t]{0.49\columnwidth}
            \centering
            \includegraphics[width=\textwidth, trim={0.6cm 0.8cm 0.4cm 0.0cm}, clip]{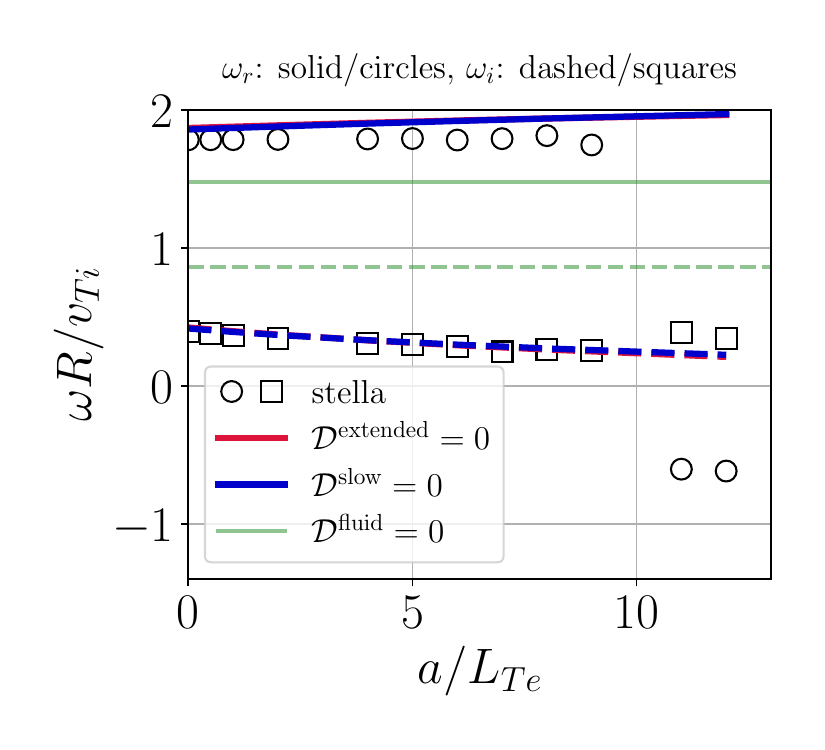}
        \caption{Electron temperature gradient scan}
        \label{fig:tailmode_omega_teprim}
    \end{subfigure}
    \begin{subfigure}[t]{0.49\columnwidth}
            \centering
            \includegraphics[width=\textwidth, trim={0.6cm 0.8cm 0.4cm 0.0cm}, clip]{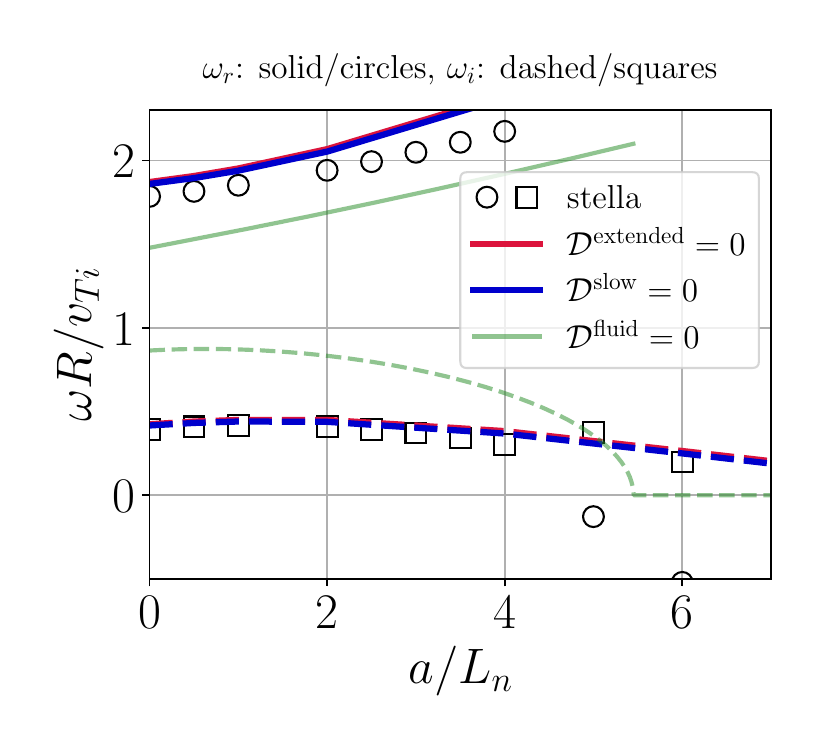}
        \caption{Density gradient scan}
        \label{fig:tailmode_omega_fprim}
    \end{subfigure}

    \caption[Extended mode frequency for varying electron temperature and density gradients]{Complex mode frequency for varying electron temperature gradient (a) and density gradient (b) -- other parameters are identical to the reference case of Figure~\ref{fig:longtail_phi_theta_ref_case}. The solid and dashed curves correspond to the real and imaginary frequencies, respectively, obtained from solving the dispersion relations \eqref{eq:tail_modes_disp_integral_eq} (red), \eqref{eq:longtail_disp_rel_weak_Pe} (blue), and \eqref{eq:longtail_disp_rel_weak_Pe_NR_ions_approximate} (green). The circles and squares correspond to the real and imaginary frequencies, respectively, obtained from gyrokinetic simulations.}
    \label{fig:tailmode_omega_teprim_fprim}
\end{figure}

Finally, we study the dependence of the extended mode frequency on the electron temperature gradient $a/L_{Te}$ and the density gradient $a/L_n$ in Figure~\ref{fig:tailmode_omega_teprim_fprim}. The extended modes become subdominant only for large values of $a/L_{Te}$ or $a/L_n$. For $a/L_{Te} \gtrsim 10$ or $a/L_n \gtrsim 5$, the fastest growing modes in gyrokinetic simulations propagate in the electron diamagnetic direction ($\omega_r < 0$). These modes are not captured by our theory; we leave their study for future work.

\section{Physics of the geodesic extended modes}\label{sec:longtail_discussion}

Based on the theory of Section~\ref{sec:longtail_theory} and the numerical insights of Section~\ref{sec:longtail_sims}, we now seek to develop physical intuition for the geodesic extended modes. Although the simulations considered only tokamak geometries, we will aim to extract from these simulations a more general understanding of GEMs relevant to both tokamak and stellarator plasmas. Instead of discussing the physics in terms of the major radius $R$ and the safety factor $q$, we will therefore use the connection length $L_\parallel$ and the magnetic curvature length scale $L_B$, with $L_\parallel = qR$ and $L_B =R $ in a tokamak.

\subsection{Rapid oscillation and connection with geodesic acoustic mode physics}\label{sec:longtail_physics_nonresonant}

The real frequency of GEMs is on the order of $\omega_r \sim v_{Ti}/L_B$, similar to the frequency of geodesic acoustic modes (GAMs) \citep{winsor_geodesic_1968, conway_geodesic_2021}. This is not coincidental, as may be understood by considering the fluid dispersion relation derived in Section~\ref{sec:longtail_model_disp}. Indeed, equation \eqref{eq:longtail_disp_rel_weak_Pe_NR_ions_approximate} has two contributions: the first, proportional to $\overline{b}_{i0}$, corresponds to the polarisation of ion gyro-orbits; the second, proportional to $\overline{\omega}_{Mi}^2$, corresponds to the geodesic curvature contribution. These two terms give rise to GAMs, whose frequency (in the absence of ion parallel streaming) may indeed be recovered exactly from \eqref{eq:longtail_disp_rel_weak_Pe_NR_ions_approximate} by considering the limit of $\omega_{*i} \sim \omega_{*i}^T \ll \omega$, such that
\begin{equation}
    0 = \mathcal{D}^\mathrm{fluid} \quad  \Leftrightarrow \quad \omega^2 = \omega_\mathrm{GAM}^2 = \frac{\overline{\omega}_{Mi0}^2}{2\overline{b}_{i0}} \left(\frac{7}{4}+\frac{1}{\tau}\right).
\end{equation}
We note that $\omega_\mathrm{GAM} \sim v_{Ti}/L_B$ and that $\omega_{*i},\omega_{*i}^T\rightarrow 0$ as $k_y \rho_i \rightarrow 0$. The GAM-like physics of the GEMs thus explains the finite GEM frequency $\omega_r \sim v_{Ti}/L_B$ as $k_y \rho_i \rightarrow 0$ (see Figures~\ref{fig:tailmode_omega_shat_small_large} and \ref{fig:tailmode_omega_ky}) and distinguishes them from localised ITG modes, whose frequency $\omega \propto k_y$ vanishes as $k_y \rho_i \rightarrow 0$.

\begin{figure}
    \centering
    \includegraphics[width=0.7\textwidth]{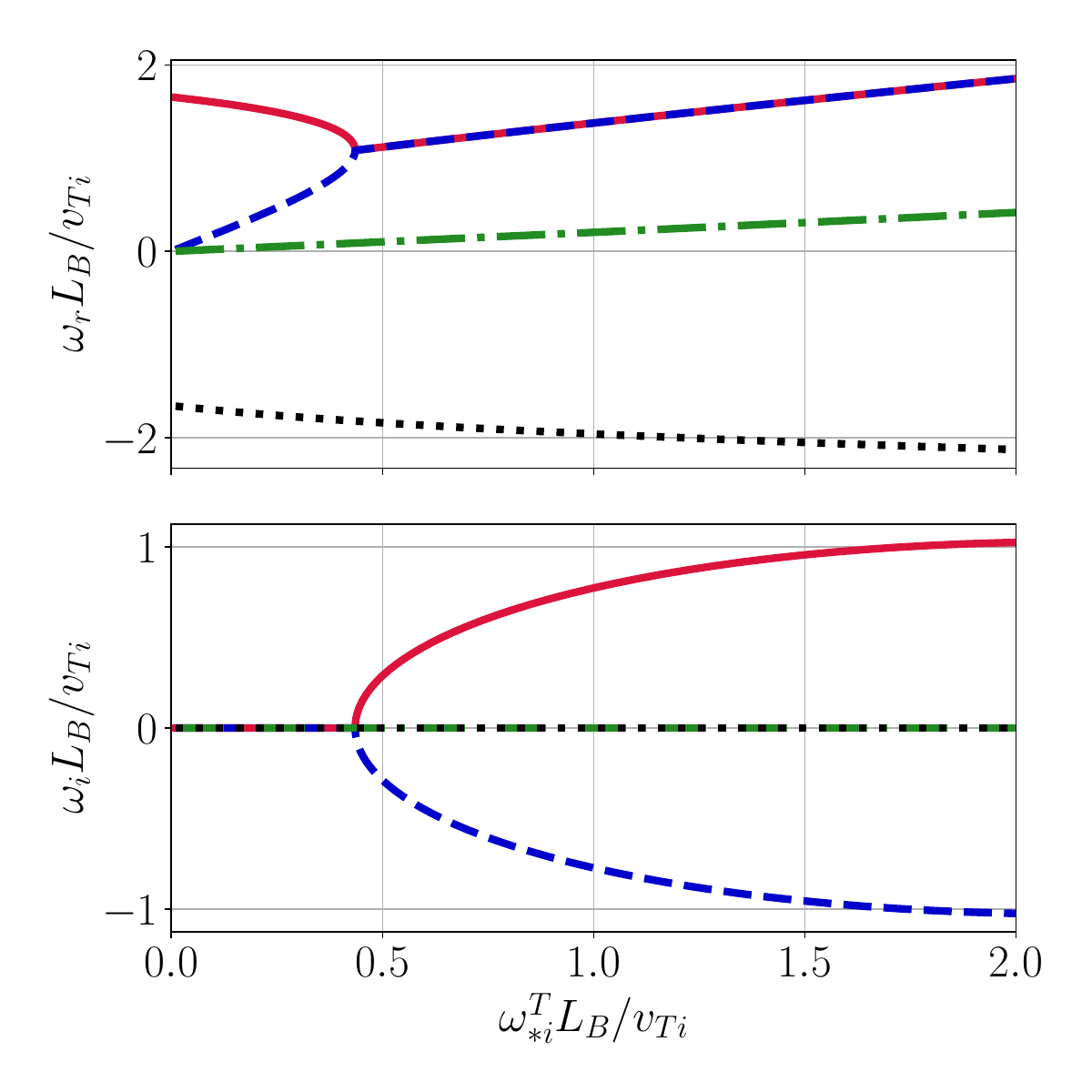}
    \caption[Branches of extended modes from fluid dispersion relation]{Real frequency (top) and growth rate (bottom) from the fluid dispersion relation \eqref{eq:longtail_disp_rel_weak_Pe_NR_ions_approximate} for a pure temperature gradient driven case ($\omega_{*i}=0$) with $\overline{b}_{i0}/\overline{\omega}^2_{Mi0} = L_B^2/v_{Ti}^2$, and electron-to-ion temperature ratio $\tau=1$. Different branches of \eqref{eq:longtail_disp_rel_weak_Pe_NR_ions_approximate} are indicated by varying colors and line styles.}
    \label{fig:extended_mode_fluid}
\end{figure}

When the ion diamagnetic frequencies $\omega_{*i}$ and $\omega_{*i}^T$ are finite, the dispersion relation \eqref{eq:longtail_disp_rel_weak_Pe_NR_ions_approximate} allows for instability, as the mode can extract free energy from the background gradients. We show in Figure~\ref{fig:extended_mode_fluid} that the fluid dispersion relation \eqref{eq:longtail_disp_rel_weak_Pe_NR_ions_approximate} describes how the mode propagating in the ion diamagnetic direction ($\omega_r > 0$) becomes unstable above a threshold in $\omega_{*i}^T$. For the zero density gradient case $a/L_n = 0$ considered in Figure~\ref{fig:extended_mode_fluid}, this threshold is at $\omega_{*i}^T L_B/v_{Ti} = k_y \rho_i L_B/L_{Ti} \approx 0.4$. This explains the GEM instability thresholds in $k_y $ and $L_B/L_{Ti}$ observed in Figure \ref{fig:tailmode_omega_ky_tiprim}. We note that, in comparison, the ITG temperature gradient threshold is $L_B/L_{Ti} \gtrsim 4$ for typical tokamak parameters \citep[see e.g.][]{dimits_comparisons_2000}.

The instability is mediated by the magnetic drifts, which aligns with the numerical result of \cite{volcokas_numerical_2024} who found very extended modes that were driven unstable by the curvature and $\nabla B$ drifts. We note that the instability mechanism is different from that of toroidal ITG modes in the sense that the GEMs are not driven by `bad curvature': the stability of the toroidal ITG depends on the combination $\omega_{*i}^T \omega_{Mi}$, with instability only when the magnetic field curvature and temperature gradient are aligned. This is not required for the GEMs, as may be seen from the fluid dispersion relation \eqref{eq:longtail_disp_rel_weak_Pe_NR_ions_approximate}.

\subsection{Vanishing adiabatic electron response} \label{sec:longtail_vanishing_electron_adiab_response}

The discussion of Section~\ref{sec:longtail_physics_nonresonant} assumed that the extended mode is entirely determined by ion physics. For this assumption to hold, the (stabilising) electron response must be negligibly small. We derived the electron density fluctuation \eqref{eq:delta_ne_total_Pe_weak} in the limit of slow electron propagation: this equation shows that there is only an adiabatic electron response on short distances $\theta_f$, as the first term in \eqref{eq:delta_ne_total_Pe_weak} vanishes upon averaging over $\theta_f$, i.e.
\begin{equation} \label{eq:delta_ne_Pe_weak_avg} 
    \frac{\langle \delta \hat n_e \rangle_{\theta_f}}{n_e} =   \frac{\omega_{*e}}{\omega} \frac{ e \langle \delta\hat\varphi \rangle_{\theta_f}}{T_e} + \frac{e}{T_e} \dd{^2 \langle \delta\hat\varphi \rangle_{\theta_f}}{\theta_s^2} \frac{\omega_{\parallel e}^2}{2\omega^2} \left( 1- \frac{\omega_{*e}+\omega_{*e}^T}{\omega} \right).
\end{equation}
Physically, the extended mode width is so large that electrons cannot respond to the potential fluctuation within the characteristic oscillation time of the mode. 

This behaviour is reminiscent of zonal flow and GAM physics, where the electrons cannot respond to radial variations of the potential as they are tied to flux surfaces and have small gyroradii. In the zonal flow case, this leads to a modified adiabatic electron response \citep{hammett_developments_1993, dorlandGyrofluidTurbulenceModels1993} with the electron density fluctuations obeying
\begin{equation} \label{eq:modified_adiabatic_zonal}
    \frac{\delta n_e}{n_e} = \frac{e}{T_e} \left( \delta\varphi - \langle \delta\varphi\rangle_\psi \right),
\end{equation}
where $\langle \rangle_\psi$ is a flux-surface average. The GAM-like physics of the GEMs is explained by the similarity of equations \eqref{eq:modified_adiabatic_zonal} and \eqref{eq:delta_ne_total_Pe_weak}, which reduces to leading order to
\begin{equation}
    \frac{\delta n_e}{n_e} = \frac{e}{T_e} \left( \delta\varphi - \langle \delta\varphi\rangle_{\theta_f} \right).
\end{equation}

The first remaining term in \eqref{eq:delta_ne_Pe_weak_avg}, proportional to $\omega_{*e}/\omega$, corresponds to the $\bs{E}\times\bs{B}$ mixing of the background electron density. In the quasineutrality equation, this term will be largely cancelled by a corresponding ion contribution (see Section~\ref{sec:longtail_model_disp}), as the $\bs{E}\times\bs{B}$ mixing of the quasineutral background does not cause charge imbalances.

The second remaining term in \eqref{eq:delta_ne_Pe_weak_avg} stems from parallel compressibility of the parallel electron flow, \textit{viz.} $\partial_t \delta n_e \sim -n_e \nabla_\parallel \delta u_{\parallel e}$. This parallel electron flow is driven by the parallel electric field and by the parallel pressure gradient, \textit{viz.} $\partial_t \delta u_{\parallel e} \sim -e \nabla_\parallel \delta\varphi/m_e -\nabla_\parallel \delta p_e /m_en_e$, with pressure fluctuations originating from mixing of the background pressure, leading to the $\omega_{*e}$ and $\omega_{*e}^T$ terms in the last parenthesis of \eqref{eq:delta_ne_Pe_weak_avg}. These terms introduce a slab electron temperature gradient (ETG) destabilisation mechanism. However, in practice, the electron response is stabilising for the GEMs, as seen e.g. in Figure~\ref{fig:tailmode_omega_teprim}, presumably due to the propagation of the GEM in the ion diamagnetic direction. The stabilising effect of electron parallel streaming aligns with the numerical results of \cite{volcokas_numerical_2024}.

\subsection{Conditions on extended mode width} \label{sec:longtail_conditions_mode_width}

A small adiabatic electron response requires the mode width to be sufficiently large. The characteristic length scale of parallel electron propagation within a mode period is given by
\begin{equation}
    \theta_\mathrm{el} \sim \frac{\omega_{\parallel e}}{\omega} \sim \frac{v_{Te}/L_\parallel}{v_{Ti}/L_B}.
\end{equation}
However, the mode cannot be infinitely extended, as it is eventually sheared to small radial scales and stabilised by ion finite Larmor radius effects. The FLR effects become important for mode widths on the order of
\begin{equation}\label{eq:longtail_thetaFLR_estimate}
    \theta_\mathrm{FLR} \sim \frac{1}{k_y \rho_i \hat s},
\end{equation}
i.e. when the mode has been sheared to the point that the radial wavelength is on the order of the ion gyroradius, $k_\perp (\theta_\mathrm{FLR})\rho_i \sim 1$. To obtain \eqref{eq:longtail_thetaFLR_estimate} from $k_\perp (\theta_\mathrm{FLR})\rho_i \sim 1$, we have used \eqref{eq:kperp_tail} and assumed small binormal wavenumbers $k_y \rho_i \ll 1$. 

The extended mode width $\Delta$ must therefore satisfy
\begin{equation} \label{eq:longtail_condition_Delta}
    \theta_\mathrm{el} \lesssim \Delta \lesssim \theta_\mathrm{FLR}.
\end{equation}
We may use this inequality to derive a condition on the size of the magnetic shear which allows for the existence of GEMs,
\begin{equation}\label{eq:longtail_condition_shat}
    \hat s \lesssim \frac{1}{k_y \rho_i} \frac{v_{Ti}/L_B}{v_{Te}/L_\parallel} \sim \frac{L_\parallel}{L_{Ti}} \sqrt{\frac{m_e}{m_i}},
\end{equation}
where the latter approximation assumes $T_e \sim T_i$ and $k_y \rho_i L_B/L_{Ti} \sim 1$ to satisfy the instability criterion (see Section~\ref{sec:longtail_physics_nonresonant}).

\begin{figure}
    \centering
    \includegraphics[width=0.6\textwidth]{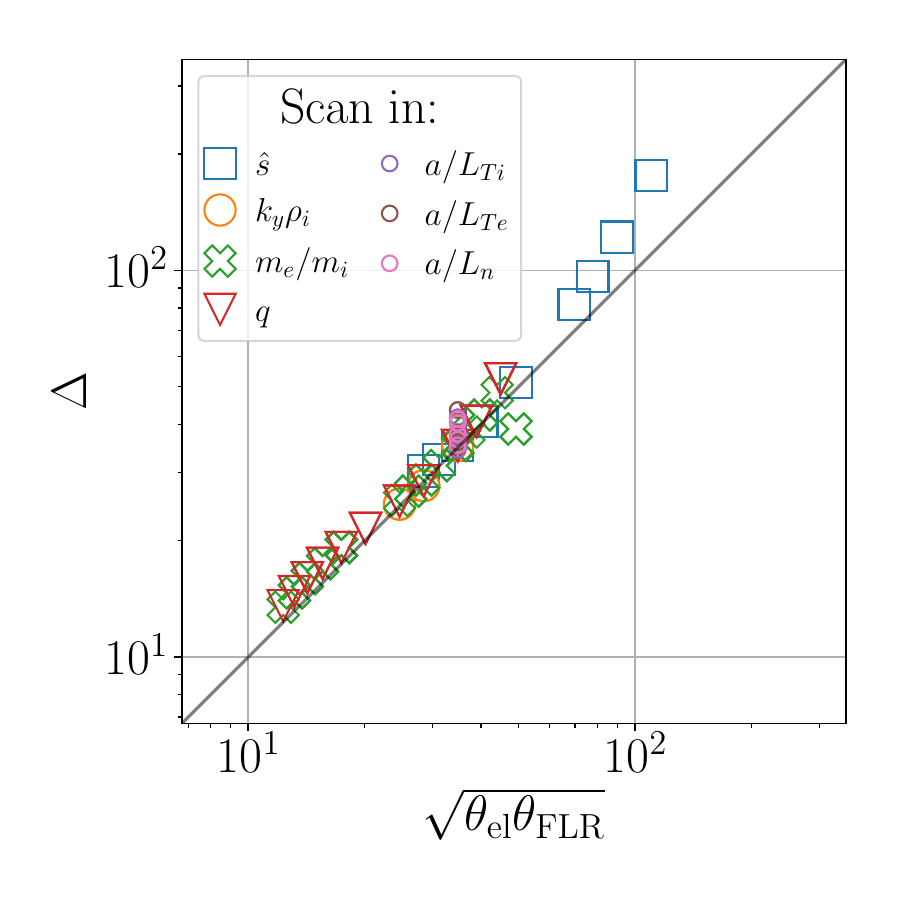}
    \caption[Extended mode width along the magnetic field]{Extended mode width $\Delta$ along the magnetic field evaluated from the gyrokinetic simulations in Section~\ref{sec:longtail_sims} using \eqref{eq:longtail_Delta}. The obtained mode width is compared with the geometric mean of $\theta_\mathrm{el} = q^{-1} (m_e/m_i)^{-1/2}$ and $\theta_\mathrm{FLR}=(k_y\rho_i\hat s)^{-1}$. To exclude gyrokinetic simulations where the localised modes are dominant, we only include those simulations where the complex mode frequency is within $10\%$ of that obtained from the dispersion relation \eqref{eq:tail_modes_disp_integral_eq}, i.e. $\abs{1-\omega_\mathrm{sim}/\omega_\mathrm{theory}} < 0.1$.}
    \label{fig:longtail_Delta}
\end{figure}

We estimate the mode width $\Delta$ in gyrokinetic simulations from the following integral over the flux-tube domain,
\begin{equation} \label{eq:longtail_Delta}
    \Delta = \left( \int\mathrm{d}\theta\, \abs{\delta\hat\varphi} \right)^{-1} \int\mathrm{d}\theta\, \abs{\theta} \abs{\delta\hat\varphi} .
\end{equation}
In Figure~\ref{fig:longtail_Delta}, we show that the mode width in gyrokinetic simulations is well approximated by $\Delta \sim \sqrt{\theta_\mathrm{el} \theta_\mathrm{FLR}}$, the geometric mean of $\theta_\mathrm{el}$ and $\theta_\mathrm{FLR}$. As $\theta_\mathrm{el} \lesssim \theta_\mathrm{FLR}$ in the region of instability, the modes do indeed satisfy the condition \eqref{eq:longtail_condition_Delta}. We note that, while $\Delta \sim \sqrt{\theta_\mathrm{el} \theta_\mathrm{FLR}}$ seems reasonably well satisfied, other scalings such as $\Delta \sim \theta_\mathrm{el}^{1/3}\theta_\mathrm{FLR}^{2/3}$ cannot be excluded based on our data. A more complete analytical theory of the dispersion relation \eqref{eq:longtail_disp_rel_weak_Pe} may shed light on the scaling of the mode width; we leave this task for future work.

Because of the condition \eqref{eq:longtail_condition_Delta}, the slow electron propagation limit of Section~\ref{sec:longtail_theory_weak_limit} is appropriate to describe the extended modes. This explains the success of the dispersion relation \eqref{eq:longtail_disp_rel_weak_Pe} in capturing the GEM frequency and eigenmode, as demonstrated by the simulations of Section~\ref{sec:longtail_sims}.

Condition \eqref{eq:longtail_condition_shat} explains why the extended modes are stabilised at large $\hat s$, small $m_e/m_i$, small $q$, and large $k_y \rho_i$, as observed in Figures~\ref{fig:tailmode_omega_shat}, \ref{fig:tailmode_omega_me}, \ref{fig:tailmode_omega_qinp}, and \ref{fig:tailmode_omega_ky}, respectively. In these limits, the electron parallel streaming is fast over the scale of the mode and the electrons are therefore adiabatic to leading order, such that only localised modes are unstable.

We note that condition \eqref{eq:longtail_condition_shat} is necessary but not sufficient. Indeed, in the limit of vanishing magnetic shear, $\hat s = 0$, we expect GEMs to exist only on flux surfaces with safety factor sufficiently close to a rational number because fluctuations extended along the magnetic field may destructively interfere on irrational flux surfaces. In flux tube simulations, this destructive interference may be modelled by introducing a phase shift in the parallel boundary condition \citep{st-onge_phase-shift-periodic_2022, volcokas_ultra_2023}. If the connection length $\theta_\mathrm{conn}$ is short compared with the typical electron propagation distance within a GEM oscillation time, $\theta_\mathrm{conn} \lesssim \omega_{\parallel e}/\abs{\omega}$, the electrons respond adiabatically and the GEMs should be stabilised. Conversely, rational flux surfaces or irrational flux surfaces with long connection lengths $\theta_\mathrm{conn} \gtrsim \omega_{\parallel e}/\abs{\omega}$ should allow for the existence of GEMs. A detailed investigation of the $\hat s = 0$ limit is outside the scope of this work.

\subsection{Limits of validity of the geodesic extended mode theory} \label{sec:longtail_limits_validity}

Equipped with our newfound understanding of the extended mode physics, we can now investigate whether the assumptions made in the derivations of Section~\ref{sec:longtail_theory} are self-consistent and how much they limit the validity of our theory.

\subsubsection{Negligible ion parallel streaming}

We neglected the parallel streaming of ions, which is justified provided that the connection length is sufficiently long,
\begin{equation}
    \frac{v_{Ti}/L_\parallel}{\omega} \sim \frac{L_B}{L_\parallel} \lesssim 1
\end{equation}
i.e. $q \gtrsim 1$ in a tokamak. This condition is similar to that for geodesic acoustic modes to avoid Landau damping. For the case of extended modes, the condition $L_\parallel \gtrsim L_B$ will only be important if it is more stringent than condition \eqref{eq:longtail_condition_Delta}, which may be rewritten as 
\begin{equation}
    L_\parallel \gtrsim L_B k_y \rho_i \hat s \frac{v_{Te}}{v_{Ti}}.
\end{equation}
Therefore, for $k_y \rho_i \hat s \,v_{Te}/v_{Ti} \lesssim 1$ and $q \sim 1$, there might be GEMs for which the ion streaming cannot be neglected. We do not consider such regimes in this work (see e.g. Figure~\ref{fig:tailmode_omega_qinp}, where the GEM growth rate vanishes for $q \lesssim 2$).

\subsubsection{Negligible averaged magnetic drifts}

In the theory of Section~\ref{sec:longtail_theory}, we assumed that $\langle \tilde\omega_{Me} \rangle_{\tau_f} = 0$, i.e. the transit-averaged electron magnetic drifts could be neglected. Using \eqref{eq:kperp_tail}, we may approximate the magnetic drift frequency as
\begin{equation}
    \tilde\omega_{M{s}} = k_y \DD{y}{{\alpha}} \vmtildea\cdot \left( \nabla\zeta\left(1 + \partial_\zeta\nu\right) + \nabla\theta \left( - q + \partial_\theta\nu \right ) - \nabla x \DD{q}{x} \theta_s  \right).
\end{equation}
The secularly increasing component due to magnetic shear only contains the radial magnetic drift, which vanishes upon a transit average over the fast coordinate for passing particles, ${\langle \vmtildea\cdot\nabla x  \rangle_{\tau_f} = 0}$. Therefore, the transit-averaged magnetic drift frequency
\begin{equation} \label{eq:longtail_avg_omegaMalpha}
    \langle \tilde\omega_{M{s}} \rangle_{\tau_f} = k_y \DD{y}{{\alpha}} \left\langle \vmtildea\cdot \left( \nabla\zeta\left(1 + \partial_\zeta\nu\right) + \nabla\theta \left( - q + \partial_\theta\nu \right ) \right) \right\rangle_{\tau_f}
\end{equation}
only has contributions from the binormal magnetic drift. At long binormal wavelengths $k_y \rho_i \ll 1$, the frequency \eqref{eq:longtail_avg_omegaMalpha} is small compared with the mode frequency $\omega \sim v_{Ti}/R$,
\begin{equation} \label{eq:small_ky}
    \frac{\langle \tilde\omega_{M{s}} \rangle_{\tau_f}}{\omega} \sim k_y \rho_i \ll 1.
\end{equation}
The condition \eqref{eq:small_ky} is sufficient for the transit-averaged electron magnetic drift to be negligible in the derivation of the electron propagator. 

For the ions, no assumptions were made about the transit-averaged magnetic drift in the derivation of the dispersion relations \eqref{eq:tail_modes_disp_integral_eq} and \eqref{eq:longtail_disp_rel_weak_Pe}, as only the local ion response needed to be evaluated, see Section~\ref{sec:longtail_theory_ions}. However, when deriving the fluid dispersion relation in Section~\ref{sec:longtail_model_disp}, we assumed that the transit-averaged magnetic drift was negligible compared to the ion polarisation contribution to quasineutrality, i.e. we assumed
\begin{equation} \label{eq:longtail_condition_Delta_omegaMi}
     \frac{\langle \tilde\omega_{Mi} \rangle_{\tau_f}}{\omega} \ll k_\perp^2 \rho_i^2 \sim (k_y \rho_i \hat s \Delta)^2,
\end{equation}
which could in theory impose a more stringent condition than \eqref{eq:longtail_condition_Delta} on the minimum mode width. We note that, in practice, the transit average (over the fast coordinate) of the binormal magnetic drift is smaller than the binormal magnetic drift itself in many cases, e.g. in a large aspect ratio tokamak $\langle \tilde\omega_{Mi} \rangle_{\tau_f} \sim (r/R) k_y \rho_i v_{Ti}/R$, which alleviates the stringency of the conditions \eqref{eq:small_ky} and \eqref{eq:longtail_condition_Delta_omegaMi}.

\subsubsection{Negligible trapped particle effects}

\begin{figure}
    \centering
            \includegraphics[width=0.8\textwidth, trim={0.6cm 0.8cm 0.4cm 0.2cm}, clip]{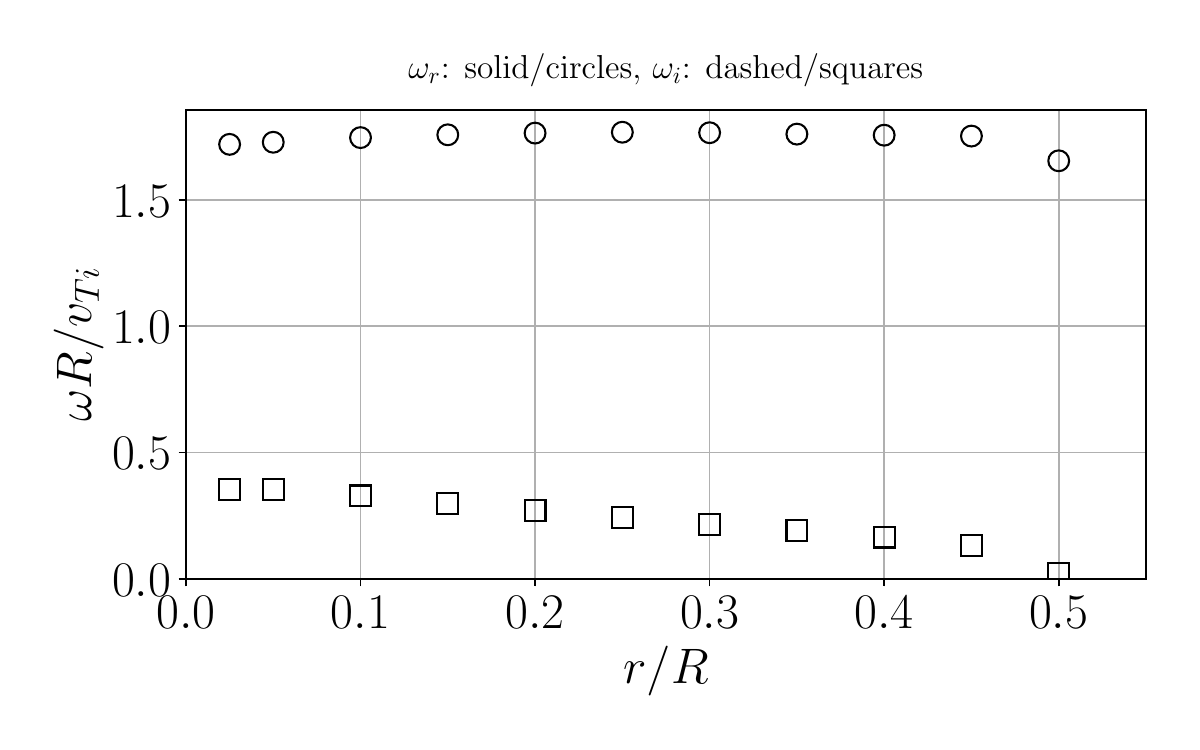}
    \caption[Extended mode frequency for varying tokamak aspect ratio]{Complex frequency for varying tokamak inverse aspect ratio $r/R$ calculated using the \texttt{stella} code. The growth rate of geodesic extended modes decreases with increasing inverse aspect ratio, but the modes are only fully stabilised for large $r/R \gtrsim 0.5$.}
    \label{fig:tailmode_omega_rhoc}
\end{figure}

The importance of trapped particle effects depends on the magnitude of the variation of $B$ along the field line. We consider numerically in Figure~\ref{fig:tailmode_omega_rhoc} how the mode frequency obtained from gyrokinetic simulations of the GEMs varies with aspect ratio. For these simulations, we consider an idealised tokamak with $R/a = 1$ and vary the flux-surface $r/a$ under consideration. All parameters are identical to the reference case of Figure~\ref{fig:longtail_phi_theta_ref_case} except for the ion temperature gradient, which is set to $a/L_{Ti}=24$ so that $R/L_{Ti}$ is the same as in the reference case. Figure~\ref{fig:tailmode_omega_rhoc} shows that the geodesic extended mode growth rate decreases as $r/R$ is increased, but full stabilisation requires large values $r/R \gtrsim 0.5$, which may limit the relevance of this effect to the outermost flux surfaces of spherical tokamaks. The stabilisation could be due to trapped particles or other geometric effects -- in any case, Figure~\ref{fig:tailmode_omega_rhoc} shows that the GEMs are not highly sensitive to the fraction of trapped particles in the tokamak case considered here. 

It is plausible that trapped particles will have a larger effect in stellarators. In particular, trapped particles in non-omnigeneous stellarators have a finite transit-averaged radial magnetic drift; our assumption of $\langle \tilde\omega_{M{s}} \rangle_{\tau_f}/\omega \ll 1$ (see discussion above) may thus be violated more easily in such devices, further limiting the parameter regime of GEMs. We leave a detailed study of these effects for future work.

\section{Extended modes in precise quasi-axisymmetric stellarator}\label{sec:longtail_precise_QA}

We show here that the geodesic extended modes are also relevant for stellarator configurations. We consider the precise quasi-axisymmetric configuration of \cite{landreman_magnetic_2022}, a stellarator with aspect ratio $R/a = 6$. This stellarator has a low magnetic shear, $\hat s \approx 0.022$ on the flux surface with normalised toroidal flux of $\psi/\mathrm{max}(\psi) =0.635$ considered here. We simulate the field line centered on $\zeta=0$ in the coordinates of the \texttt{VMEC} code \citep{hirshman_three-dimensional_1986} -- though we do not show this here, other field lines exhibit similar behaviour.

\subsection{Linear simulations}

First, we consider linear simulations similar to those presented for tokamaks in Section~\ref{sec:longtail_sims}. Here, we use a flux tube extending for $600$ field periods (approximately $126$ poloidal turns as the rotational transform $\iota = 1/q \approx 0.42$ and the number of field periods is two) with $3600$ grid points. The velocity-space resolution is $N_{v_\smallparallel}=48, N_\mu = 8$, max$(v_\parallel)=3 v_{Ti}$, and the maximum of $\mu$ is chosen such that the maximum perpendicular velocity is $3 v_{Ti}$ at the minimum value of $B$. We consider a large ion temperature gradient $a/L_{Ti}=4$, a moderate density gradient $a/L_n=1$, and no electron temperature gradient, $a/L_{Te}=0$. 

\begin{figure}
    \begin{subfigure}[t]{0.49\columnwidth}
    \centering
    \includegraphics[width=\textwidth, trim={0.6cm 0.8cm 0.4cm 0.0cm}, clip]{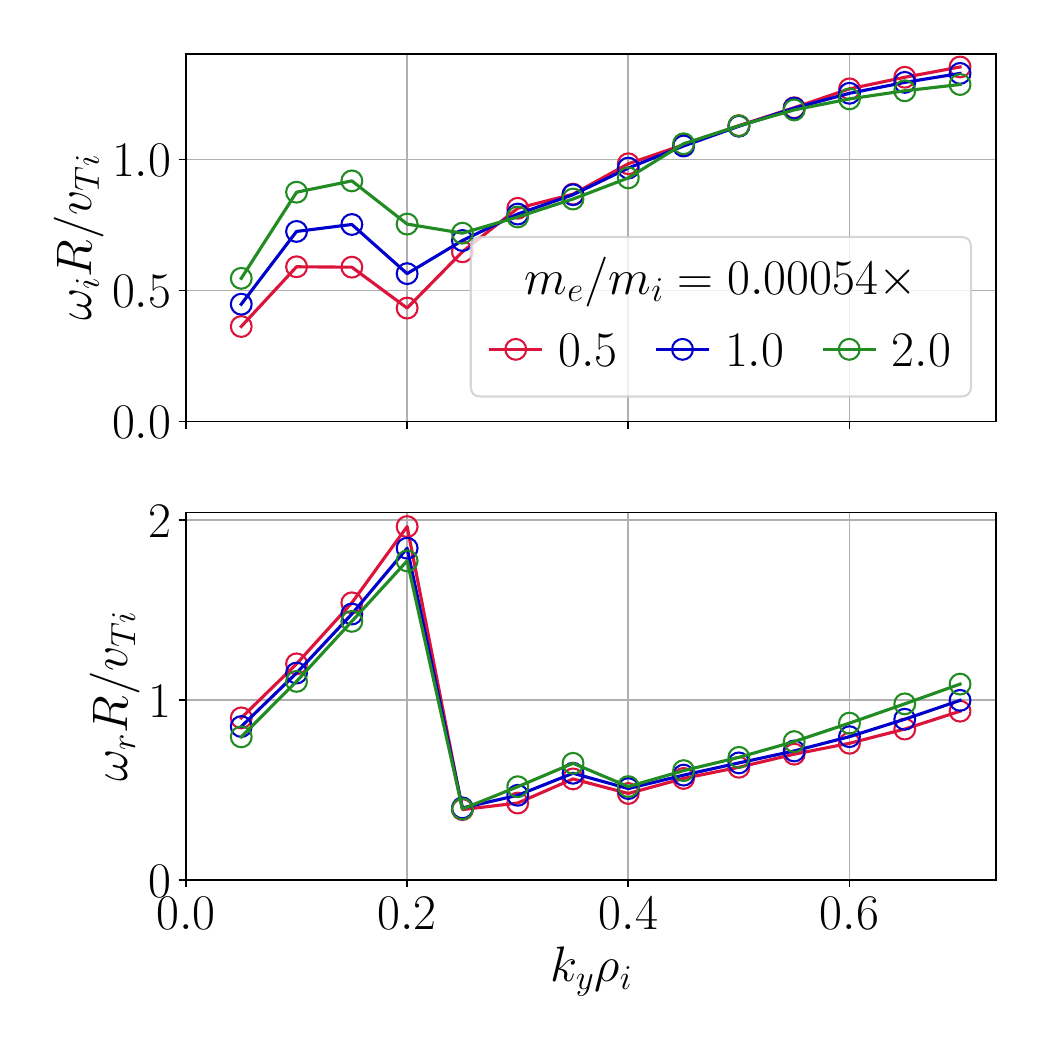}
    \caption{Growth rate (top) and real frequency (bottom) as a function of the binormal wavenumber $k_y \rho_i$ for different values of the electron mass (colors).}
    \label{fig:longtail_omega_precise_QA_omega}
    \end{subfigure}
    \begin{subfigure}[t]{0.49\columnwidth}
    \centering
    \includegraphics[width=\textwidth, trim={0.6cm 0.8cm 0.4cm 0.0cm}, clip]{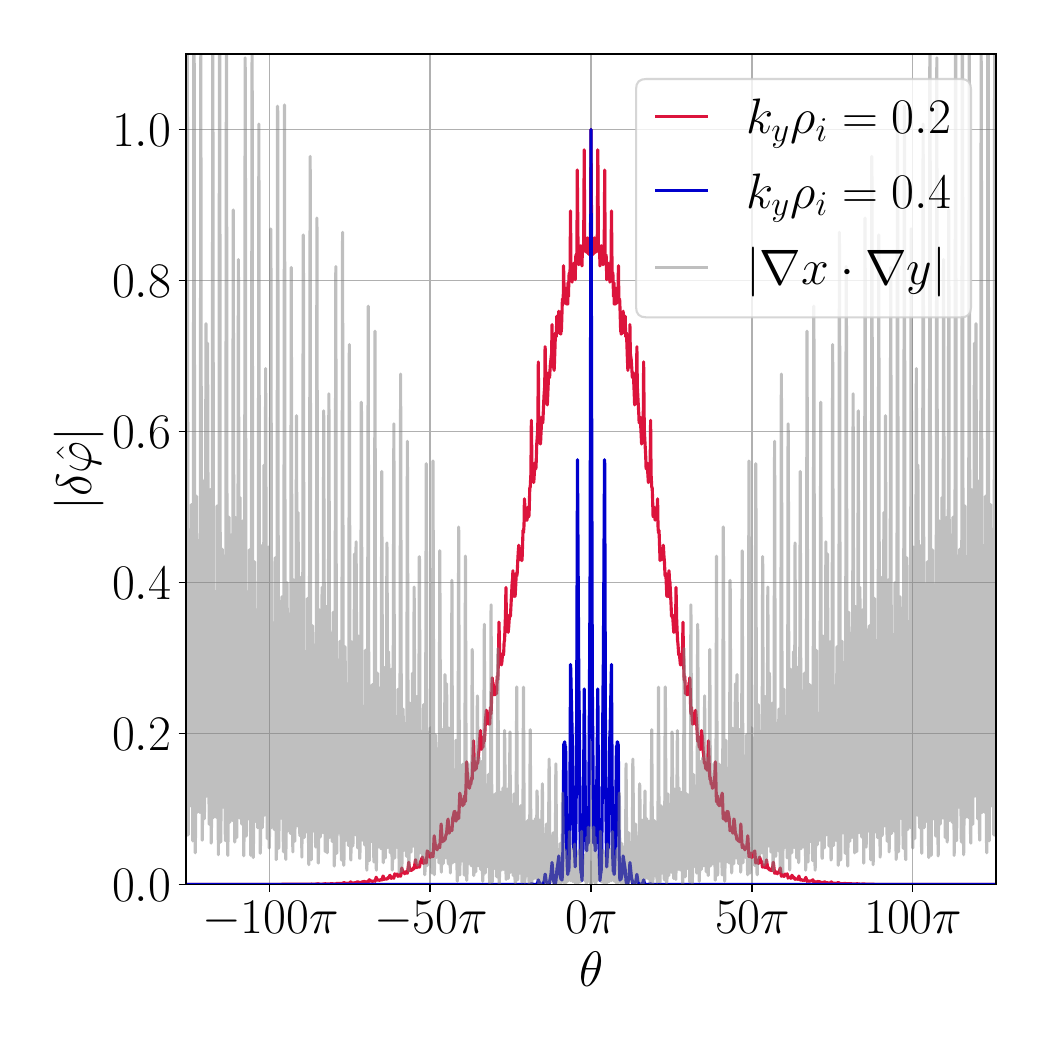}
    \caption{Potential fluctuation ballooning structure along the magnetic field for ${m_e/m_i = 0.00054}$ and two values of the binormal wavenumber (red and blue), and geometric coefficient $\abs{\nabla x \cdot \nabla y}$ (grey).}
    \label{fig:longtail_omega_precise_QA_phi}
    \end{subfigure}
    \caption[Extended modes in precise quasi-axisymmetric stellarator]{Complex frequency (a) and eigenmode structure (b) from linear gyrokinetic simulations in the precise quasi-axisymmetric stellarator of \cite{landreman_magnetic_2022}. The high-frequency geodesic extended modes are the fastest growing modes at small binormal wavenumbers $k_y \rho_i\lesssim 0.2$.}
    \label{fig:longtail_omega_precise_QA}
\end{figure}

We show in Figure~\ref{fig:longtail_omega_precise_QA} that the GEMs are clearly observed in the precise quasi-axisymmetric stellarator. Like in the tokamak case (see Figure~\ref{fig:tailmode_shat_small_large}), they manifest as rapidly oscillating modes at small $k_y \rho_i$ (Figure~\ref{fig:longtail_omega_precise_QA_omega}) which are extended along the magnetic field (Figure~\ref{fig:longtail_omega_precise_QA_phi}). Their growth rate increases as the electron mass is artificially increased, due to the weakening of the stabilising electron response -- see discussion in Section~\ref{sec:longtail_vanishing_electron_adiab_response}. At large $k_y \rho_i$, localised modes with smaller $\omega_r$ grow faster than the GEMs, with a growth rate that is independent of the electron mass.

We note here that, contrary to tokamaks, stellarators like the precise quasi-axisymmetric stellarator exhibit a large variation of the local magnetic shear $\nabla x \cdot \nabla y$ along the magnetic field, as shown in Figure~\ref{fig:longtail_omega_precise_QA_phi}. At least in the case considered here, these `shear spikes' do not prohibit the existence of GEMs.

\subsection{Nonlinear simulations}

\begin{figure}
    \centering
    \begin{subfigure}[t]{\columnwidth}
            \centering
            \includegraphics[width=\textwidth, trim={2cm 0.2cm 5cm 1.25cm}, clip]{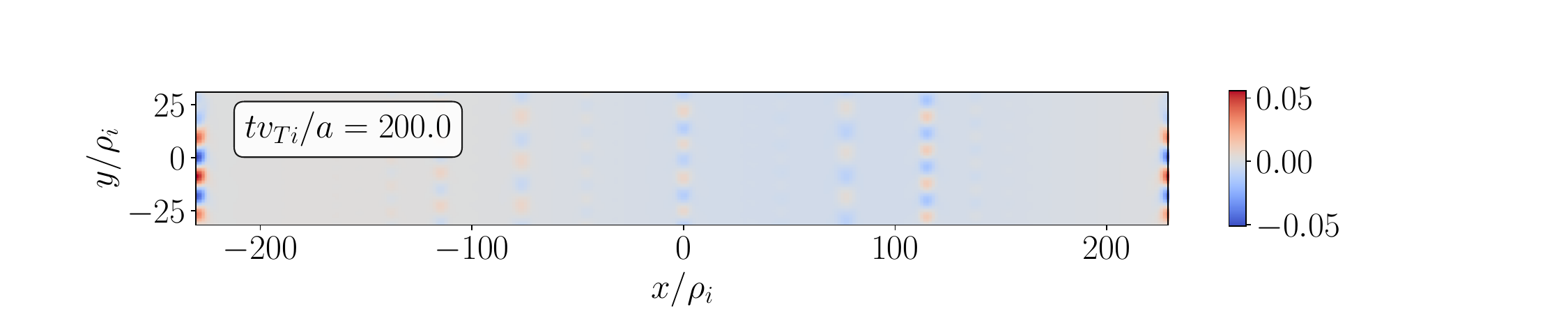}
        \caption{Linear stage: instability at low-order rationals}
    \end{subfigure}
    \begin{subfigure}[t]{\columnwidth}
            \centering
            \includegraphics[width=\textwidth, trim={2cm 0.2cm 5cm 1.25cm}, clip]{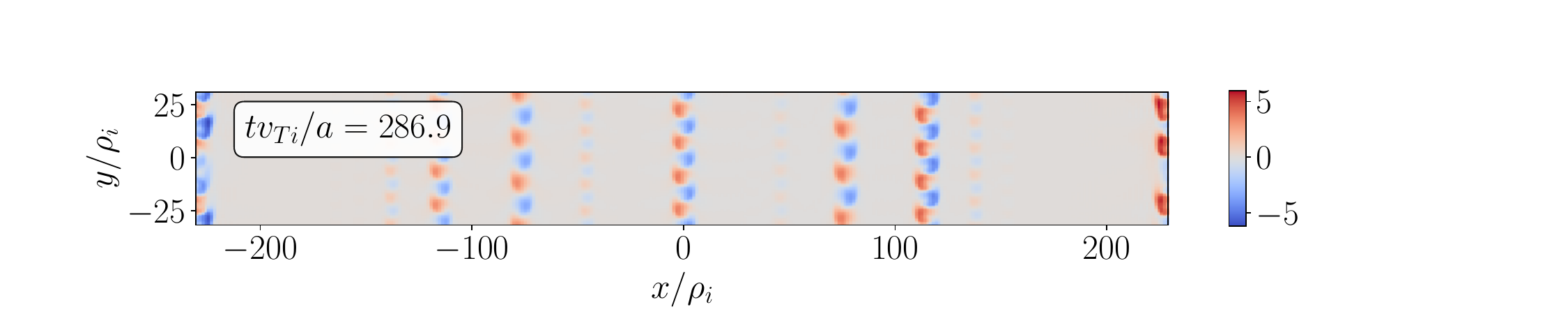}
        \caption{Onset of nonlinearity}
    \end{subfigure}
    \begin{subfigure}[t]{\columnwidth}
            \centering
            \includegraphics[width=\textwidth, trim={2cm 0.2cm 5cm 1.25cm}, clip]{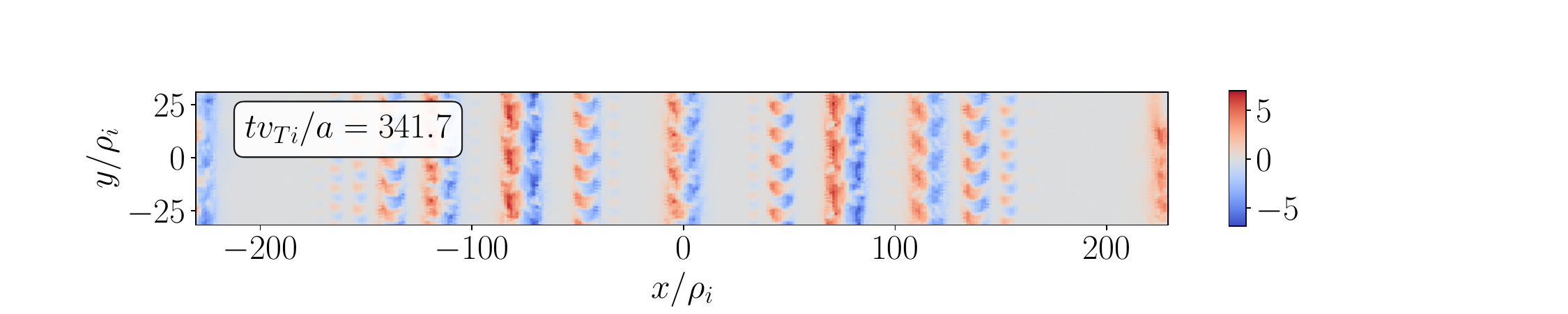}
        \caption{Nonlinear structures at low-order rationals}
    \end{subfigure}
    \begin{subfigure}[t]{\columnwidth}
            \centering
            \includegraphics[width=\textwidth, trim={2cm 0.2cm 5cm 1.25cm}, clip]{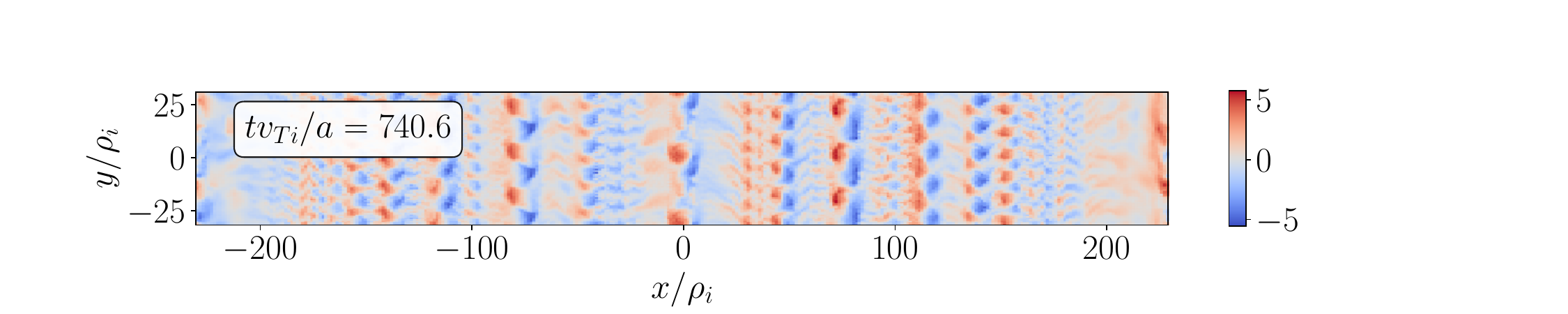}
        \caption{Nonlinear structures fill most of the radial domain}
    \end{subfigure}
    \caption[Extended modes in nonlinear simulations of precise quasi-axisymmetric stellarator]{Normalised potential fluctuations $e\delta\varphi/T_i (a/\rho_i)$ at $\theta=0$ for various times (top to bottom) in a nonlinear simulation of the precise quasi-axisymmetric stellarator of \cite{landreman_magnetic_2022}.}
    \label{fig:longtail_NL_precise_QA}
\end{figure}

We show in Figure~\ref{fig:longtail_NL_precise_QA} the real space variation of the turbulent fluctuations from a nonlinear gyrokinetic simulation of the precise quasi-axisymmetric stellarator. Here, the flux tube length is chosen to extend for $4.782$ field periods, so that the ends of the tube are at the maximum of the quasisymmetric $B$ on the surface. The parallel grid consists of $24$ grid points and the velocity-space resolution is given by $N_{v_\smallparallel}=32, N_\mu = 6$, max$(v_\parallel)=3 v_{Ti}$, and the maximum of $\mu$ is chosen such that the maximum perpendicular velocity is $3 v_{Ti}$ at the minimum value of $B$. The twist-and-shift boundary condition \citep{beerFieldalignedCoordinatesNonlinear1995} is used: due to the low shear, this sets the perpendicular domain aspect ratio to $\Delta k_x/\Delta k_y = 2\pi \hat s \approx 0.14$. The binormal resolution is $\Delta k_y \rho_i = 0.1$, the number of radial modes is $N_x = 512$, and the number of binormal modes is $N_y = 128$. The background gradients are set to $a/L_{Ti}=0.9$, $a/L_n = 0.5$, and $a/L_{Te}=0$.

The twist-and-shift boundary condition introduces mode-rational surfaces in the simulation domain, as discussed e.g. in \cite{volcokas_numerical_2024}: the flux surfaces at $x=0$ and $x=\pm \pi/\Delta k_x$ are periodic at the tube ends, the flux surfaces at $x=\pm \pi/(2\Delta k_x)$ are periodic after two tube `turns', etc. By tube `turns', we mean the traversal of the tube extent in the parallel direction and the subsequent application of the twist-and-shift boundary condition. On an irrational flux surface, the extended mode eigenfunction cannot be sustained due to destructive interference, as each tube `turn' will add an irrational phase shift. Therefore, the extended tail modes are predominantly observed at the low-order mode rational surfaces, as shown in Figure~\ref{fig:longtail_NL_precise_QA}. As time progresses, structures become visible around higher order rationals, and eventually fill the entire radial domain.

\section{Summary and discussion} \label{sec:longtail_conclusion}

In this article, we presented a theory of extended modes at low magnetic shear. This theory naturally led to the geodesic extended modes (GEMs), previously identified as a `$k_z \approx 0$' branch of the toroidal ITG modes in \cite{volcokas_ultra_2023, volcokas_numerical_2024}. The GEMs have a frequency $\omega \sim v_{Ti}/R$ (which is large compared to diamagnetic frequencies at small $k_y \rho_i$) and are very extended along the magnetic field. Indeed, the extent of these modes along the magnetic field must be sufficiently large for the electron parallel streaming term to become small despite the electrons' fast propagation speed. In the limit where the electron parallel streaming term is negligibly small, the GEMs can be described by a simple fluid dispersion relation \eqref{eq:longtail_disp_rel_weak_Pe_NR_ions_approximate}, which exhibits characteristics similar to the dispersion relation for geodesic acoustic modes, as discussed in Section~\ref{sec:longtail_physics_nonresonant}.  The stabilisation due to the finite electron parallel streaming is adequately captured by the numerical solutions of the dispersion relations \eqref{eq:tail_modes_disp_integral_eq} and \eqref{eq:longtail_disp_rel_weak_Pe}.

For the sake of simplicity, we studied the GEM physics considering mostly a simple tokamak geometry with circular cross-section. As a proof of principle, we also showed in Section~\ref{sec:longtail_precise_QA} that the GEMs are present in linear and nonlinear simulations of the precise quasi-axisymmetric stellarator \citep{landreman_magnetic_2022}.  Although only this stellarator was considered here, we anticipate GEMs to be more generally relevant for all tokamaks and stellarators with global magnetic shear sufficiently low to satisfy condition \eqref{eq:longtail_condition_shat}. For instance, the GEMs exist in the precise quasi-axisymmetric stellarator despite large variations of the local magnetic shear (see Figure~\ref{fig:longtail_omega_precise_QA_phi}), and the GEMs are only weakly sensitive to the device aspect ratio (see Figure~\ref{fig:tailmode_omega_rhoc}).

Future studies ought to consider the nonlinear GEM physics and their impact on turbulence, topics which are outside the scope of this paper. On the one hand, one may expect the GEMs to decorrelate nonlinearly more easily than localised modes due to their extent along the magnetic field lines. On the other hand, the self-interaction of extended modes has previously been shown to nonlinearly modify zonal profiles, e.g. the zonal flows \citep{weikl_occurrence_2018, c_j_how_2020} and the safety factor profile \citep{volcokas_turbulence-generated_2025}. The linear physics and the parametric dependences of the GEMs elucidated in this study may thus prove important for the description of turbulence in tokamaks and stellarators with low magnetic shear.

\vphantom{bla}\\
\noindent\textbf{Data availability statement.} The data supporting the findings of this article is openly available at \href{https://datacommons.princeton.edu/discovery/catalog/doi-10-34770-7z83-hy37}{https://datacommons.princeton.edu/discovery/catalog/doi-10-34770-7z83-hy37}.

\vphantom{bla}\\
\noindent\textbf{Funding.} This work was supported by U.S. DOE DE-AC02-09CH11466 through PPPL's Laboratory Directed Research and Development (LDRD). The simulations presented in this article were performed on computational resources managed and supported by Princeton Research Computing, a consortium of groups including the Princeton Institute for Computational Science and Engineering (PICSciE) and the Office of Information Technology's High Performance Computing Center and Visualization Laboratory at Princeton University.

\vphantom{bla}\\
\noindent\textbf{Declaration of interests.} The authors report no conflict of interest.

\appendix

\FloatBarrier

\bibliographystyle{jpp}
\bibliography{references}

\end{document}